\documentclass[twocolumn,showpacs,amsmath,amssymb,floatfix,pre]{revtex4}

\usepackage{graphicx}
\usepackage{dcolumn}
\usepackage{bm}

\begin{document}


\preprint{Franzese at al.}

\title{Metastable liquid-liquid phase transition in a single-component
system with only one crystal phase and no density anomaly} 
\author{G. Franzese$^{1,2}$} 
\altaffiliation[Present address:~SMC, Dipartimento di Fisica, Universit\`a ``La Sapienza'', P.le A. Moro 2, I-00185 Roma, Italy]{}
 \email{franzese@na.infn.it}
\author{G. Malescio$^3$}
\author{A. Skibinsky$^1$}
\author{S. V. Buldyrev$^1$}
\author{H. E. Stanley$^1$}

\affiliation{$^1$Center for Polymer Studies and Department of Physics,
Boston University, Boston, MA 02215, USA\\ 
$^2$Dipartimento di Ingegneria
dell'Informazione,   
Seconda Universit\`a di Napoli, 
Istituto Nazionale Fisica della Materia
UdR Napoli and CG SUN, 
I-81031, Aversa, Italy\\
$^3$Dipartimento di Fisica,
Universit\`a di Messina and Istituto Nazionale Fisica della Materia,
I-98166 Messina, Italy}

\date{\today}

\begin{abstract}
We investigate the phase behavior of a single-component system in
3 dimensions with spherically-symmetric, pairwise-additive,
soft-core interactions with an attractive well at a long distance, a
repulsive soft-core shoulder at an intermediate distance, and a hard-core
repulsion at a short distance, similar to potentials used to describe
liquid systems such as colloids, protein solutions, or liquid metals.
We showed [Nature {\bf 409}, 692 (2001)]
that, even with no evidences of the density anomaly,
the phase diagram has two 
first-order fluid-fluid phase transitions, one ending in a 
gas--low-density liquid (LDL) critical point, and the other in a 
gas--high-density liquid (HDL) critical point, with a LDL-HDL phase
transition at low temperatures.
Here we use integral equation calculations
to explore the 3-parameter space of the soft-core potential and
we perform molecular dynamics simulations in the
interesting region of parameters.
For the equilibrium phase diagram we analyze
the structure of the crystal phase and find that, within the considered
range of densities, the structure is independent of the density.  
Then, we analyze in detail the fluid metastable phases and,
by explicit thermodynamic calculation
in the supercooled phase, we show the absence of the density anomaly.
We suggest that this absence is related to the presence of only one
stable crystal structure.
\end{abstract}


\maketitle

\section{Introduction}

Soft-core potentials have been widely used to study a variety of systems
such as liquid metals, metallic mixtures, electrolytes and colloids, as
well as anomalous liquids, like water and silica
\cite{Debenedetti,hs7072,Silbert76_Levesque77_Kincaid78,%
Shyu71_Appapillai72_Mon79_Kahl84,Lawrence75,colloids,DRB91,%
SH-G93,Debene,ssbs,nature,pellicane,jagla,jagla2,PhysABig}.  
In these models, the
specific structural characteristic at the molecular (or atomic) level are
neglected and the molecules (or atoms) are represented by simple spheres. 
Quantum effects (such as the quantum nature of chemical interactions) and
classical effects (such as the Coulomb interaction) are modeled through a
phenomenological isotropic pair potential.  The advantages of this
approach are that while these potentials are simple enough to be treated
analytically \cite{Hansen-McDonald}, they still allow a qualitative
comparison with the experiments. Moreover, they can be studied by means of
numerical simulations less time-consuming than those of realistic models
\cite{Frenkelbook}. 

We consider an off-lattice model in three dimensions (3D) \cite{nature}
related to the soft-core potentials studied by Hemmer and Stell
\cite{hs7072} for solid-solid critical points.  Our model shows a phase
diagram with two fluid-fluid phase transitions, a feature recently seen in
experiments on phosphorus \cite{Katayama_Falconi} and confirmed by
specific simulations \cite{Morishita01}.  In Ref.~\cite{nature} we showed
that both first-order fluid-fluid phase transitions end in critical
points, a low-density critical point $C_1$, and a high-density critical
point $C_2$.  For the considered potential, both transitions occur in the
supercooled phase with respect to the crystal phase. 

\begin{figure}
\includegraphics[width=8cm,height=6.5cm,angle=0]{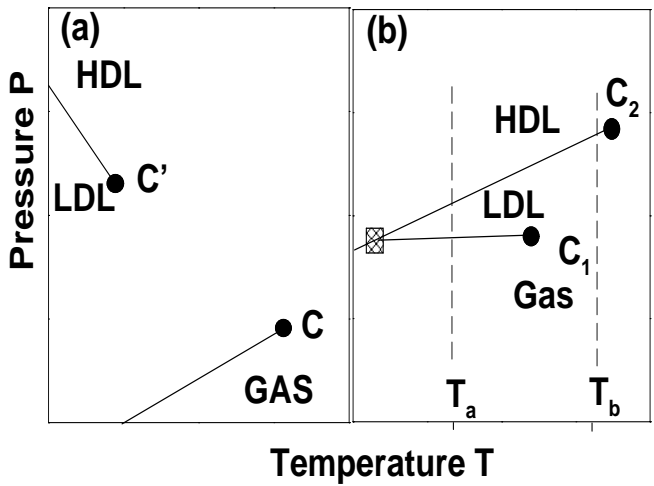}
\caption{Schematic pressure--temperature $P-T$ phase diagrams with two
critical points. Solid lines represent first order phase transition lines,
circles represents critical points. (a) Phase diagram with critical point
$C$ between the gas and the uniform liquid phase at high $T$ and low $P$
and critical point $C'$ between the low-density liquid (LDL)  and the
high-density liquid (HDL) at low $T$ and high $P$. This phase diagram has
been proposed for water and has been shown to be consistent with the
density anomaly. (b) Phase diagram with critical point $C_1$
between the gas and the LDL at low $T$ and low $P$ and the critical point
$C_2$ between the gas and the HDL at high $T$ and high $P$.  Increasing
$P$ at constant temperature $T_a$ below $C_1$ (dashed line at $T_a$), the
system undergoes a first order phase transition between the gas and the
LDL phase, followed by a first order phase transition between the LDL and
the HDL phases.  Increasing the pressure at constant temperature $T_b$
above $C_1$ but below $C_2$ (dashed line at $T_b$), the system undergoes
only a first order phase transition between the gas and the HDL. The
square represents the gas-LDL-HDL triple point. This phase diagram has
been found in Ref.~\protect\cite{nature} with no evidence of the
density anomaly.} 
\label{schem}
\end{figure}

The hypothesis of a second critical point has been proposed \cite{Poole}
as a way to rationalize the density anomaly---i.e., the expansion upon
isobaric cooling---in network-forming fluids, such as water
\cite{Debenedetti,Poole,llcp_water}, carbon \cite{carbon}, silica
\cite{silica} and silicon \cite{sastry_volga}. Consequently, both the
experimental \cite{ms98,Bellissent98,Soper00,Brazhkin,Wilding} and the
theoretical 
\cite{DRB91,SH-G93,Debene,ssbs,jagla,jagla2,PhysABig,Errington,Guisoni} 
investigations
about the possibility of a second critical point have been focused on
systems with the density anomaly (Fig.~\ref{schem}a).  However, the
results of Ref.~\cite{nature} have shown that the presence of the critical
point $C_2$ does not necessarily induce the density anomaly, indicating
that the quest for simple liquids with two critical points is not
restricted to systems with densities exhibiting anomalous behavior
(Fig.~\ref{schem}b). 
In this paper we push forward the analysis, by studying the equilibrium
phase diagram and showing that the system introduced in
Ref.~\cite{nature} has 
one single crystal phase, in the range of considered densities,
suggesting that the absence of density anomalies is related to the
presence of only one stable crystal structure. 

To reach this goal we organize the work in the following way.
After the definition of our soft-core potentials (Section II),  
(i) we describe in detail the integral equations in
the hyper-netted chain (HNC) approximation (Section III.A)
and
(ii) we generate new calculations
using different assigned parameters for the pair potential, thus
establishing a bridge between the potential studied in Ref.~\cite{nature}
and the potential investigated in Ref.~\cite{ssbs,PhysABig} and rationalizing
how---in addition to the critical point $C_1$---the critical point $C_2$
arises as a function of the parameters of the pair potential (Section III.B). 
Then, 
(iii) we describe in detail the molecular dynamics (MD) simulations,
studying the equilibrium phase diagram and
finding that the only stable crystal structure, in the range of simulated
densities, has two characteristic lattice distances (Section IV.A). 
(iv) By MD simulations we analyze also the supercooled liquid phase
and the metastable fluid-fluid phase transition (Section IV.B);
(v) in order to construct an accurate phase diagram and to study the
finite size effect, we perform new MD simulations in addition to the
calculations presented in Ref.~\cite{nature} (Section IV.C).
Hence, 
(vi) we study the radial distribution function, 
comparing the HNC predictions with the MD results and analyzing the
composition of the system within the fluid-fluid coexisting regions
(Section V).
Finally, 
(vii) we address the density anomaly
issue by presenting the explicit thermodynamic calculation, based on
the MD phase diagram, that
allows us to exclude the presence of the density anomaly (Section VI).
We give our conclusions in Section VII.

\section{Soft-core potentials}

Among the isotropic potentials, much attention has been devoted to
soft-core potentials, which have a finite (soft-core) repulsion added to
the infinite (hard-core) repulsion.  The infinite repulsion is due to the
impenetrability of the spheres.  The finite repulsion represents the
combination of all the quantum and classical repulsive effects averaged
over the angular part.  It has been shown
\cite{Shyu71_Appapillai72_Mon79_Kahl84} that a weak effective repulsion
can be derived by a first principle calculation for liquid metals
\cite{Silbert76_Levesque77_Kincaid78}.  To understand the possibility of
the solid-solid critical point in such material as Ce and Cs, Hemmer and
Stell \cite{hs7072} proposed a soft-core potential with an attractive
interaction at large distances, performing an exact analysis in 1D.  Over
the past thirty years, several other soft-core potentials with one or more
attractive wells were proposed and studied with approximate methods, or
numerical simulations in 2D, to rationalize the properties of liquid
metals, alloys, electrolytes, colloids and the water anomalies
\cite{Debenedetti,hs7072,Silbert76_Levesque77_Kincaid78,%
Shyu71_Appapillai72_Mon79_Kahl84,Lawrence75,colloids,DRB91,%
SH-G93,ssbs,nature,pellicane,jagla,jagla2,PhysABig}.  
It has been recently shown
\cite{nature}, for the first time in 3D with MD, that an appropriate
soft-core potential with an attractive well is able to show two
supercooled liquids of different densities, with two critical points.
Similar results have been reported for a 
soft-core potentials with
a linear repulsive ramp by Monte Carlo simulations \cite{jagla2}.

Following Ref.~\cite{nature}, we define the isotropic pair potential
$U(r)$, as a function of the pair distance $r$ (inset in Fig.\ref{ie1}): 
\begin{equation}
U(r)=\left\{
\begin{array}{lll}
\infty & \mbox{  for  } & r < a \\
U_R    & \mbox{  for  } & a\leq r < b \\
-U_A    & \mbox{ for  } & b\leq r <c  \\
0      & \mbox{  for  } & c\leq r ~~~ ,
\end{array}\right.
\label{Uofr}
\end{equation}
where $a$ is the hard-core distance and $b$ is the soft-core distance. 
For $a\leq r <b$, the spheres interact with a finite (soft-core) repulsive
energy $U_R$.  For $b\leq r <c$, the pair's interaction is attractive with
energy $-U_A<0$.  The distance $c$ is the cut-off radius beyond which the
pair's interaction is considered negligible.  For sake of comparison with
Ref. \cite{ssbs,PhysABig}, we will consider both $U_R>0$ and $U_R<0$.

The step-like shape in Eq.~(\ref{Uofr}) has the advantage of being defined
by only three parameters: the width of the soft-core in units of the
hard-core distance $w_R/a\equiv (b-a)/a$, the width of the attractive well
in the same units $w_A/a\equiv (c-b)/a$ and the soft-core energy in units
of the attractive well energy $U_R/U_A$.  To explore the phase diagram of
the model as a function of these three parameters in an approximate, yet
fast way, we use the integral equations in the HNC approximation. 

\section{The integral equations in the hyper-netted chain approximation}

In this section we present  details of the integral equations and
the HNC approximation adopted in Ref.~\cite{nature} and in this work.  The
radial distribution function $g(r)$ plays a central role in the physics of
fluids \cite{Hansen-McDonald}.  This quantity is proportional to the
probability of finding a particle at a distance $r$ from a reference one
and is the ratio of local to bulk density at distance $r$, with 
\begin{equation}
g(r) \equiv 
\frac{n(r)}{4\pi r^2 \rho} ~,
\label{g_r}
\end{equation}
where $n(r)$ is the number of particles at a distance between
$r$ and $r+dr$ from the reference particle and $\rho$ is the number
density, assumed to be independent of $r$ (uniform system).
The radial
distribution function goes to 1 for large $r$ and is always 1 for a random
spatial distribution of particles.  To represent deviations from
randomness, the {\em total} pair correlation function $h(r)\equiv g(r)-1$
is introduced. 

These functions are relevant because they are directly measurable by
radiation scattering experiments and are related to the thermodynamic
properties of the fluid. A fundamental relation between structure and
thermodynamics is given by
\begin{equation}
k_B T\rho K_T=1+\rho\int h(\vec{r})~ d\vec{r} ~,
\label{K_T}
\end{equation}
where $K_T\equiv (\partial \rho/\partial P)_T /\rho$ is the isothermal
compressibility, $P$ is the pressure, and $k_B$ is the Boltzmann constant. 
Provided that the particles interact through pairwise-additive forces,
other thermodynamic properties of the fluid---such as the internal
energy---can be expressed using integrals over the pair correlation
function. 

The function $h(r)$ is the result of the interaction of all the particles
in the system. Formally, $h(r)$ can be decomposed into (i) the
contribution coming from the {\em direct} interaction between two
particles at distance $r$, called $c(r)$, and (ii) the contribution due to
the {\em indirect} interaction propagated through any other particle in
the system. This second contribution is written in turn as an integral
convolution of direct correlations and total pair correlation. 

This decomposition, for uniform systems, is expressed by the
Ornstein-Zernike relation
\begin {equation}
h(r)=c(r)+\rho \int c(r') h(|\vec{r}-\vec{r'}|) \; d\vec{r'} ~ .
\label{OZ}
\end {equation}
Equation~(\ref{OZ}) is also the formal definition of $c(r)$.  Both $h(r)$
and $c(r)$ in Eq.~(\ref{OZ}) are unknown functions, thus to solve this
equation, one needs another relation ({\em closure}) between these two
functions.  This relation is provided by the diagrammatic expansion of
$g(r)$ \cite{Hansen-McDonald} which, after formal summation, yields the
functional relation
\begin {equation}
g(r)=\exp [- \beta U(r) + h(r) - c(r)+d(r)] ~ ,
\label{closure}
\end {equation}
where $U(r)$ is the inter-particle potential, $\beta\equiv 1 /(k_B T)$ and
$d(r)$ is the sum over a specific class of diagrams ({\it bridge
diagrams}) \cite{Hansen-McDonald}.  Since $d(r)$ cannot be calculated
exactly, one resorts to approximate expressions.  The simplest
approximation assumes $d(r)=0$ (HNC closure) \cite{HNC1}.  One expects
this approximation to work better at lower $\rho$, where the direct
correlation function $c(r)$ is more relevant then the correlation
propagated through the other particles. However, our results (see
Section~V) will show that this intuitive observation is not
straightforward, at least for soft-core potentials. 

\subsection{The iterative procedure}

The solution of the integral Eqs.~(\ref{OZ},\ref{closure}) with the
HNC closure is obtained through a numerical iterative procedure whose
essential scheme is the following. Under the assumption $d(r)=0$, one can
write $g(r)$ as
\begin {equation}
g(r)=\exp [- \beta U(r) + \theta (r)] ~ ,
\label{gofr}
\end {equation}
where the function $\theta (r) \equiv h(r)-c(r)$ has the remarkable
property of being a continuous function of $r$, even for discontinuous
potentials (as in this paper).  From the definitions of $h(r)$,
$\theta(r)$ and Eq.~(\ref{gofr}), one can derive the equation
\begin {equation}
c(r)=\exp [- \beta U(r) + \theta (r)] - \theta (r) - 1 ~.
\label{cofr}
\end {equation}
By using the Fourier transform $\hat{f}(\vec{q})\equiv \int f(\vec{r}) 
\exp(i\vec{q}\cdot \vec{r}) ~d\vec{r}$ defined for a generic function
$f(\vec{r})$, from Eq.~(\ref{OZ}) we obtain
\begin {equation}
\hat{h}(q) = \hat{c}(q) + \rho~ \hat{c}(q)~ \hat{h}(q) ~.
\end {equation}
Or, using the definition of $\theta(r)$, we have
\begin {equation}
\hat{\theta} (q) = \rho~ \frac{\hat{c}^2(q)}{1-\rho~ \hat{c}(q)} ~
\label{thetaofq}
\end {equation}
The numerical iteration is based on Eqs.~(\ref{cofr}) and
(\ref{thetaofq}). 

We start by choosing an initial guess for $\theta (r)$.
A reasonable input, at least at high temperatures,  is the $\theta (r)$ of
a fluid of hard spheres
with diameter $a$. In fact, at high temperatures our potential can be
approximated with a
simple hard-core repulsion. We can calculate the corresponding $\theta
(r)$ by making use of
the Percus-Yevick integral equation \cite{Hansen-McDonald} 
which for hard spheres can be solved analytically.
Next, at constant $\rho$, we decrease the temperature of $\delta T$ and
we perform calculations
at fixed $\rho$ and $T$ by using as input the $\theta (r)$ obtained as
solution at $\rho$ and $T+\delta T$.

From the chosen guess of $\theta (r)$ we calculate $c(r)$ by using
Eq.~(\ref{cofr}).
Its Fourier transform $\hat{c}(q)$ is used in
Eq.~(\ref{thetaofq}) to calculate $\hat{\theta} (q)$. Its inverse Fourier
transform provides a new $\theta (r)$ that is used as a new input for the
next cycle.  We evaluate the functions on $M=2048$ discrete points
$r_m=m\delta r$, with $m=1,\dots, M$ and $\delta r=0.01 a$.  Successive
iterations of the elementary cycle defines a succession
${\theta^{(k)}(r)}$, where $k=1, 2, \dots$ is the number of the iteration.
If the difference between two consecutive elements of this succession,
\begin {equation}
\Delta \equiv \left[\frac{1}{M}\sum_{m}^{M}
[\theta^{(k+1)}(r_m) - \theta^{(k)}(r_m)]^2 \right]^{1/2},
\end {equation}
decreases for increasing $k$, the succession converges towards a
$\theta^\ast(r)$ that is solution of our integral equations.  The
iteration process is stopped when $\Delta\leq 10^{-7}$.

Based on this iterative procedure, different algorithms can be used to
improve the accuracy and rapidity of convergence of the numerical solution
of HNC equations. However, independent of the algorithm used, there exists
a region in the $\rho$-$T$ plane where no solution can be found, i.e., for
any $\rho$, there is a $T$ below which the numerical algorithm does not
converge, defining an {\it instability line} in the $\rho$-$T$ plane. 

\subsection{The HNC instability line}

The nature of the locus of instabilities of the HNC equation and 
its relationship with the spinodal line of the fluid was investigated 
for a hard-core potential plus an attractive Yukawa tail in a number 
of papers \cite{FA,PA}.
These studies showed that the isothermal compressibility does not 
diverge as the temperature is lowered and the instability region is 
approached from above. 
This conclusion was definitively assessed through extensive 
numerical calculations \cite{B}
both for the hard-core Yukawa fluid and 
other model potentials, showing that this
behavior is
directly correlated to the existence of multiple HNC solutions. 
The analysis developed was based on a careful 
treatment of the low-$k$ behavior of the Fourier transforms of the 
correlation functions required by the iterative procedure.
A further theoretical support to these results was given by an
analysis \cite{H} on models for an ionic fluid and a monoatomic
Lennard-Jones fluid.

In the light of the above mentioned studies, an identification of the 
instability line of the HNC equation with the spinodal line of the fluid, 
which is characterized by a diverging compressibility, is not possible.  
Keeping in mind this limitation, one can nevertheless observe that 
for a large number of simple fluid pair potentials the shape of the 
instability line qualitatively resembles the region of spinodal decomposition 
of the fluid.
Also for our potential the comparison of the HNC calculations with the 
MD results (Section~\ref{HNCvsMD}) shows that the HNC
instability line is qualitatively consistent with the spinodal line
\cite{notaHNC}.  
Thus, studying the modifications of the instability line as the 
potential parameters are varied can yield some approximate, yet useful, 
informations on the phase behavior of the fluid.

\subsection{The results}

\begin{figure}
\includegraphics[width=8cm,height=6.5cm,angle=0]{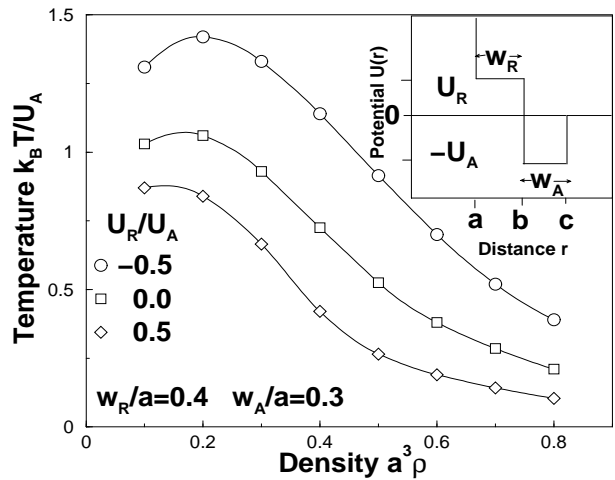}
\caption{Instability line of the HNC equations in 3D in the $\rho$-$T$
plane for the pair potential in Eq.~(\protect\ref{Uofr}), with $w_R/a=0.4$,
$w_A/a=0.3$ and (from top to bottom) $U_R/U_A=-0.5$, 0.0, 0.5.  For
$U_R/U_A=-0.5$, the pair potential recovers the one studied in 1D and 2D
in Ref.~\protect\cite{ssbs,PhysABig}.  
The symbols represent the calculations and
the lines are guides for the eyes.
Inset: the isotropic pairwise-additive potential $U(r)$ as a function of
the inter-particle distance $r$; $a$ is the hard-core distance, $b$ is the
soft-core distance, $c$ is the maximum attractive distance, $-U_A<0$ is
the attractive energy and $U_R$ is the repulsive energy.} 
\label{ie1} 
\end{figure}

First, we calculate the instability line of the HNC equations for the
potential investigated in Refs.~\cite{ssbs,PhysABig}. The
corresponding parameters are $w_R/a=0.4$, $w_A/a=0.3$ and $U_R/U_A=-0.5$. 
In this case, the soft-core is given by two attractive wells with
different depths.  Calculations in 1D and 2D \cite{ssbs,PhysABig} have shown a
water-like density anomaly. Therefore, it is interesting to analyze the
phase diagram in 3D.  However, the instability line for this case
(Fig.~\ref{ie1}) is similar to the spinodal line usually exhibited by a
simple fluid, e.g. interacting via a Lennard-Jones potential with the
maximum of the spinodal line corresponding to the liquid-gas critical
point.  Upon increasing $U_R/U_A$ to 0.5 (Fig.~\ref{ie1}), the only
evident change of the instability line is a shift toward a lower $T$ as a
result of the overall decrease of the inter-particle attraction, with no
hints of a second critical point. A small shift to lower $\rho$ is also
seen. This behavior is more evident for larger $w_R$. 

\begin{figure}
\includegraphics[width=8cm,height=6.5cm,angle=0]{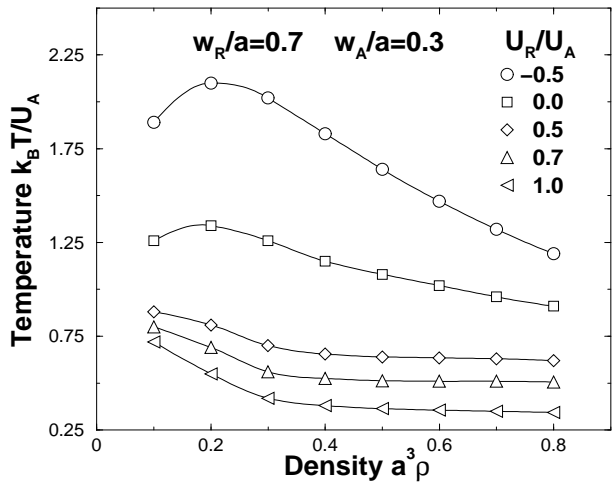}
\caption{Instability lines, as in Fig.~\protect\ref{ie1}, for the pair
potentials with parameters $w_R/a=0.7$, $w_A/a=0.3$ and (from top to bottom) 
$U_R/U_A=-0.5$, 0.0, 0.5, 0.7, 1.0.} 
\label{ie2} 
\end{figure}

Next, we consider a potential with a larger $w_R$ ($w_R/a=0.7$,
$w_A/a=0.3$).  The instability line is calculated for several values of
$U_R/U_A$ (Fig.~\ref{ie2}). Upon increasing $U_R/U_A$, we now find not
only the shift to a lower $T$, but also an evident shift of the maximum of
the line (i.e. the critical point, 
assuming that the instability line represents the behavior
of the spinodal line) to a lower $\rho$. This result can be rationalized
by observing that, passing from $U_R<0$ to $U_R>0$, the soft-core becomes
more and more difficult to penetrate and the system passes from a
potential with a hard-core $a$ and an effective attractive range $w_A+w_R$
to a potential with an effective hard-core $b$, for $U_R/U_A$ large
enough, and an attractive range $w_A$.  As a consequence of the increase
of the effective hard-core, the critical density decreases and, as a
consequence of the decrease of the effective overall attraction, the
critical temperature decreases. 

Comparing Fig.~\ref{ie1} and Fig.~\ref{ie2}, we notice an important
difference.  In the case of larger $w_R$ (Fig.~\ref{ie2}), as $U_R/U_A$
increases, the temperature of the instability line does not decrease for
increasing $\rho$, but becomes rather flat.  This result suggests that the
instability line might develop a second maximum at larger values of $\rho$
for even larger $w_R$. 

\begin{figure}
\includegraphics[width=8cm,height=6.5cm,angle=0]{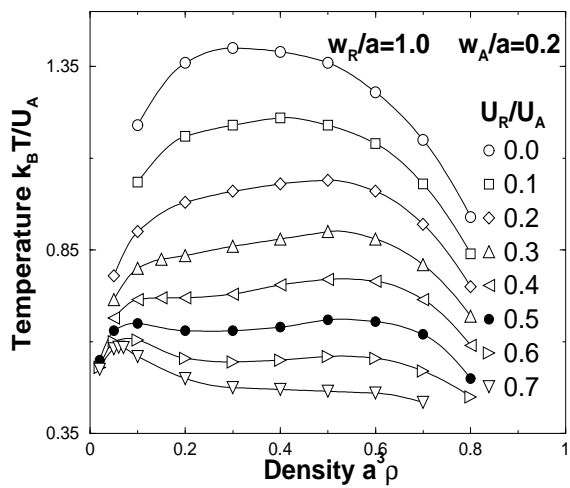}
\caption{Instability lines, as in Fig.~\protect\ref{ie1}, for the pair
potentials with parameters $w_R/a=1.0$, $w_A/a=0.2$ and (from top to bottom)
$U_R/U_A=0.0$, 0.1, 0.2, 0.3, 0.4, 0.5, 0.6, 0.7. The full symbols
correspond to the set of parameters selected for the MD calculations.}
\label{ie3} 
\end{figure}

We thus consider a potential with $w_R/a=1.0$ and $w_A/a=0.2$.  The
results (Fig.~\ref{ie3}) show that for $0.4\leq U_R/U_A \leq 0.6$, the
instability line has two well-distinct local maxima, suggesting the
possibility of two critical points in the phase diagram for the fluid
phases \cite{nota}.  For $U_R/U_A\leq 0.3$ or $U_R/U_A\geq 0.7$, the
instability line shows just one maximum, similar to the typical spinodal
line of a fluid of hard spheres with diameter $a$ or $b$, respectively,
attracting via a square well of width $w_A$.  As a consequence of this
analysis, we choose $w_R/a=1$, $w_A/a=0.2$ and $U_R/U_A=0.5$ as the set of
parameters for the potential used in the MD calculations in 3D
\cite{nature}. 

\section{The molecular dynamics approach} 

In this Section we give extensive details on the MD method and we extend
the analysis performed in Ref.\cite{nature}, including new
calculations for the crystal phase, the crystal nucleation process and
the metastable phases.
We perform MD simulations at constant number of particles $N$ of unit mass
$m$, at constant volume $V$, with periodic boundary conditions and at
constant average temperature $T$.  We present the results for $N=490$,
720 and
$N=1728$.  The average temperature is set by coupling the system to a
thermal bath at the assigned $T$, with a thermal exchange coefficient per
particle between the system and the bath equal to $k=0.015~
(U_A/m)^{1/2}k_B/a$.  We use a standard collision event list algorithm
\cite{r} to evolve the system and a modified Berendsen method to achieve
the desired $T$ \cite{b}. 

The pressure is calculated by using the virial expression for a step
potential \cite{Frenkelbook}
\begin{equation}
P=\frac{m}{3V}\left\langle \sum_i^N v_i^2 +
\frac{1}{\Delta t} 
\sum_{i,j}^N \hspace{-.05cm}' \vec{\Delta v}_i \cdot
(\vec{r}_i-\vec{r}_j)\right\rangle ~ ,
\end{equation}
with ${\sum'}_{i,j}^N$ sum over the particles pairs $(i,j)$ undergoing a
collision in the time interval $\Delta t \equiv (10^5
ma^2/U_A)^{1/2}$, hereafter used as unit of time, and
with $\vec{\Delta v}_i \equiv \vec{v}_i'-\vec{v}_i$, where $\vec{v}_i$ and
$\vec{v}_i'$ are the velocities of the particle $i$ at position
$\vec{r}_i$ before and after the collision with particle $j$ at position
$\vec{r}_j$. 

\subsection{The crystal}

First, to locate the equilibrium crystal line we simulate a crystal seed
surrounded by the gas.  We prepare a crystal seed by
cooling at $T=0.45U_A/k_B$ a gas configuration with density
$\rho=0.018/a^3$.

\begin{figure}
\includegraphics[width=8cm,height=8cm,angle=0]{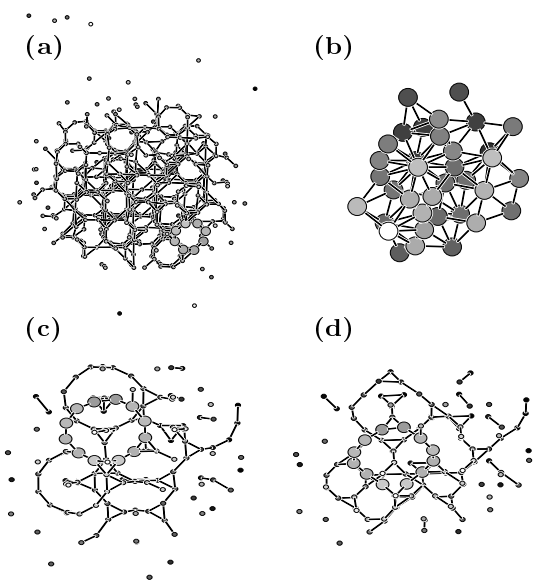}
\caption{The MD configuration equilibrated 
at $k_BT/U_A=0.45$ and $a^3\rho=0.018$.
Darker particles are farther away from the observation point. 
(a) The crystal, with defects, surrounded by the gas. 
Bonds connect particles at distance $1\leq r/a \leq 1.2$.  
The radius of the particles is {\it not} in scale with the distances. 
A typical ring of 8 particles (octagon) is plotted with a larger radius
(b) A section of the crystal.  
Bonds connect particles at distance $2\leq r/a \leq 2.2$. 
The radius of the particles is in scale
with the distances.
Note, in the upper part of the panel, a ring of 12 nn particles  
(dodecagon), connected by bonds to two central particles.
(c) The same 12 particles of the section in (b) 
are plotted with a larger radius with
respect to the other particles. Bonds are like in panel (a).
(d) A rotation of 40 degrees around a central horizontal axis of the section
in (c) reveals the 8-fold symmetry observed in (a).} 
\label{xtal} 
\end{figure}

The crystal (Fig.~\ref{xtal}) is the effect of the competition between the
hard-core repulsion at distance $r=a$ and the attraction at distance
$r=b$.
The resulting structure
is reminiscent of the close packing of hard spheres with diameter $a$ or
$b$, but the competition gives rise to new symmetries (Fig.~\ref{xtal}). 
The minimum in the inter-particle interaction potential at $b\leq r<c$
would induce a face-centered-cubic crystal with lattice space ranging from
$b$ to $c$ and a characteristic six-fold symmetry on one projection plane
with seven particles to form a triangular lattice.  However, due to the
soft-core, the system can allocate particles at the hard-core distance
$a=b/2$.  This induces a twelve-fold symmetry, placing an average number
of twelve, almost on-plane, nearest neighbor (nn) particles at a distance
$1\leq r/a\lesssim 1.2$ to form a dodecagon around two particles.  These
two nn particles are next nearest neighbors 
to the dodecagon, at a
distance $2\leq r/a \leq 2.2$, and are placed on a line almost
perpendicular to the plane individuated by the dodecagon
(Fig.~\ref{xtal}b).  This structure is distorted in such a way to form
non-closed chains of nn particles, that wrap along another axis to give
rise to an eight-fold symmetry (Fig.~\ref{xtal}c,d). 

\begin{figure}
\Large{\bf (a)}\includegraphics[width=3cm,height=3cm,angle=0]{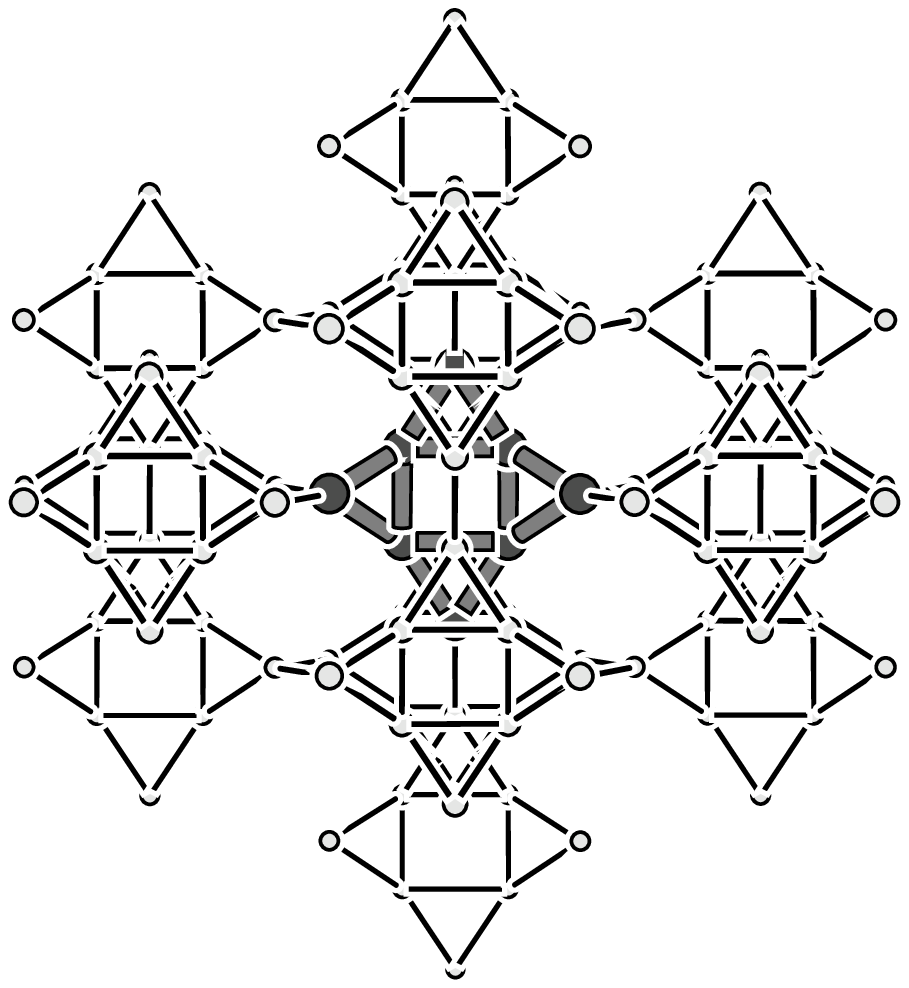} 
\Large{\bf (b)}\includegraphics[width=3cm,height=3cm,angle=0]{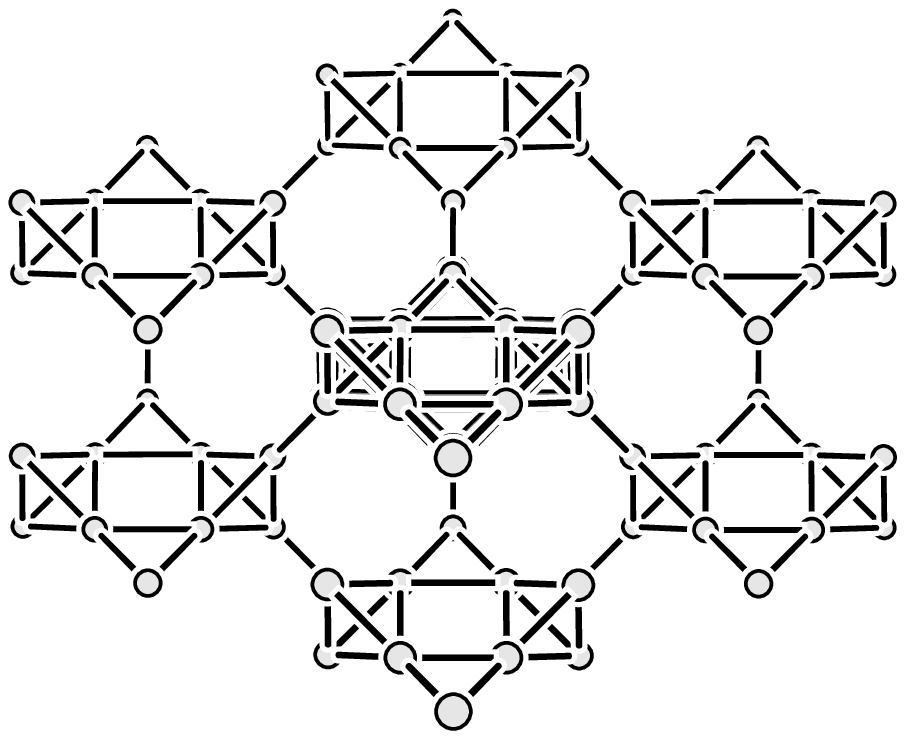} 
\Large{\bf (c)}\includegraphics[width=3cm,height=3cm,angle=0]{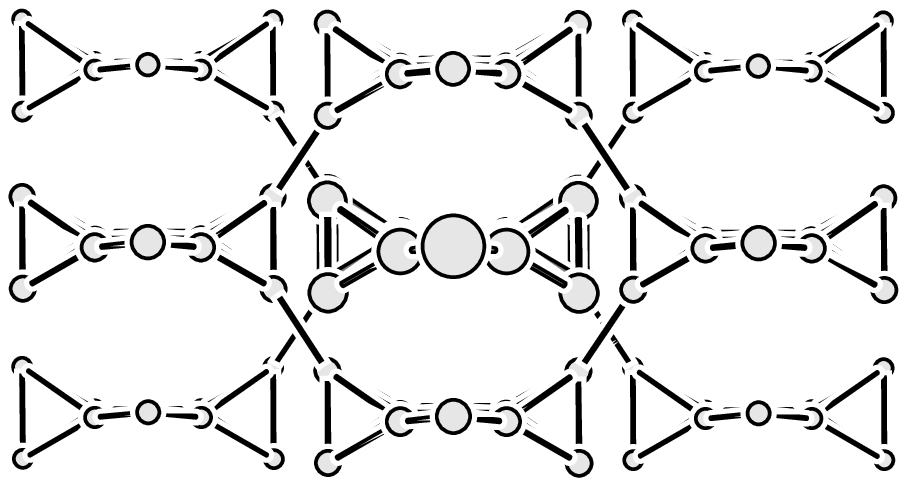} 
\Large{\bf (d)}\includegraphics[width=3cm,height=3cm,angle=0]{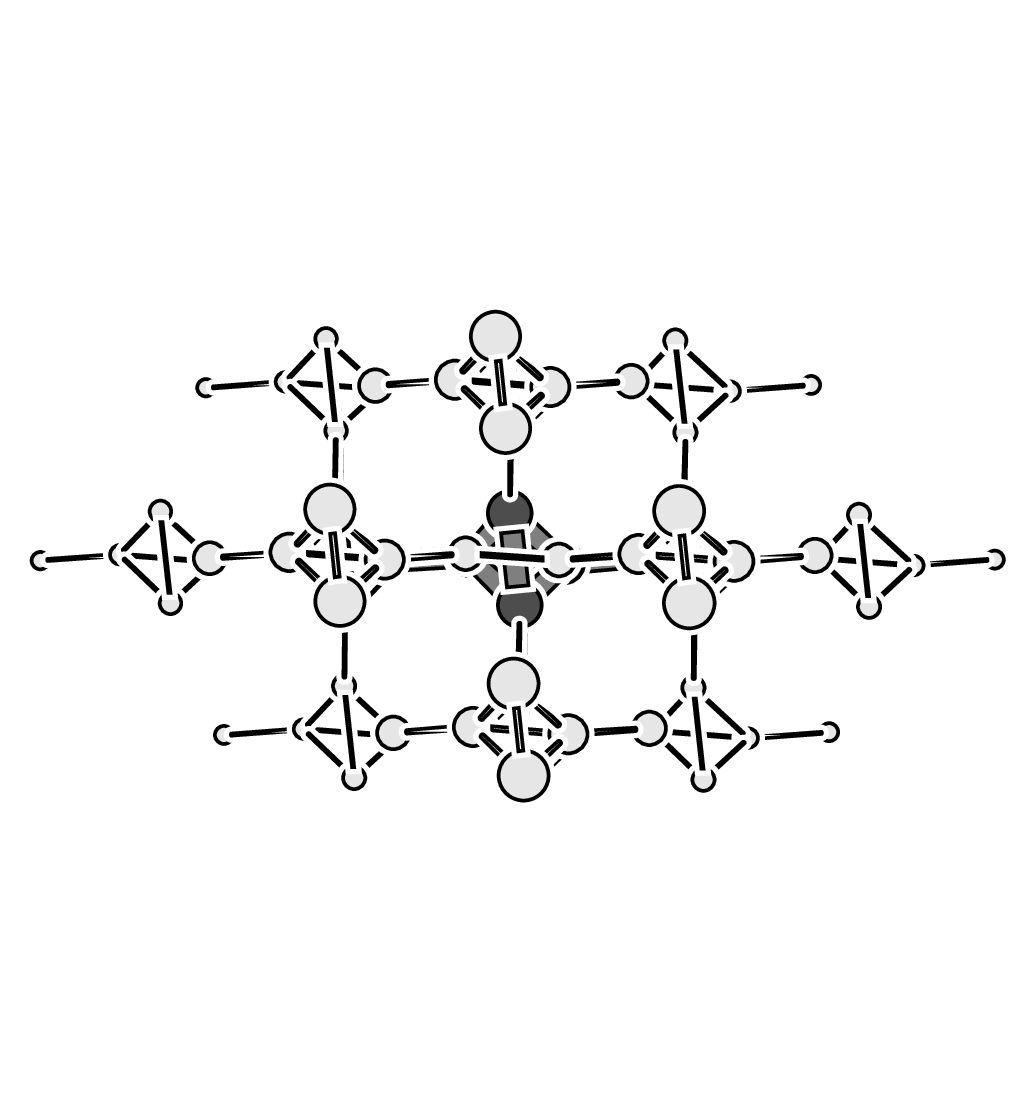} 
\caption{The artificial crystal configuration equilibrated for a time
$10^2 \Delta t$ with a MD simulation at $T=0.03U_A/k_B$. Bonds
connect particles at distance $r/a\leq 1.2$. The radius of the
particles is {\it not} in scale with the
distances. Greater particles are closer to the observation point. The
configuration contains 15 cells. The central cell is emphasized by
darker bonds. 
(a) Each cell has 4 particles at the corners of a rectangle 
($\vec{r}_m$ with $m=1, \dots 4$ in Table~\protect\ref{tab3}), whose
long sides form two triangles with two particles on the same plane 
($\vec{r}_m$ with $m=5, 6$ in Table~\protect\ref{tab3}), and the  
short sides forms two tetrahedra, each with two more particles 
($\vec{r}_m$ with $m=7, \dots 10$ in Table~\protect\ref{tab3}).
(b) The crystal configuration rotated by $\pi/4$ around a central
horizontal axis shows the 8-fold symmetry seen in
Fig.~\protect\ref{xtal}a and d. 
(c) A further rotation of $\pi/4$ around the same axis shows the
dodecagons seen in Fig.~\protect\ref{xtal}b and c. 
(d) A rotation of $\pi/2$ around a central vertical axis shows again
the octagons seen in Fig.~\protect\ref{xtal}a and d. 
} 
\label{artificial} 
\end{figure}

By analyzing the crystal structure obtained from the MD simulations,
we conclude that the position of the particles in the crystal can be
described by $\vec{r}=i\cdot \vec{a}+j\cdot \vec{b}+k\cdot \vec{c}+
\vec{r}_m$, where $\vec{r}_m$, for $m=1, \dots 10$, are the
coordinates, with respect to the center of the cell, of the 10
particles forming a crystal cell, 
$\vec{a}$, $\vec{b}$ and $\vec{c}$ are the lattice vectors describing the
position of the center of the cell and $i$, $j$ and $k$ are integers
such that $i+j+k$ is even.
We estimate the lattice vectors (Table~\ref{tab2})
and the coordinates of the particles forming the cell (Table~\ref{tab3})
after an equilibration time $10^2 \Delta t$ 
at $T=0.03U_A/k_B$ for of an
artificial crystal placed in the vacuum (Fig.~\ref{artificial}). 
The resulting density of the crystal is $a^3\rho\simeq 0.39$.

\begin{table}  
\caption{The coordinates of the lattice vectors 
as obtained after an equilibration time $10^2 \Delta t$ 
at $T=0.03 U_A/k_B$ of the artificial crystal
(Fig.~\protect\ref{artificial}) proposed 
to describe the crystal structure found in the MD simulations
(Fig.~\protect\ref{xtal}). The errors on the parameters are on the
last decimal digit and decrease as the square root of the time of
averaging.}    
\begin{ruledtabular}
\begin{tabular}{c | c c c }  
            & $x$    & $y$    & $z$    \\ 
\hline
$\vec{a}/a$ & $1.95$ & $0.00$ & $0.11$ \\
$\vec{b}/a$ & $0.00$ & $3.41$ & $0.00$ \\ 
$\vec{c}/a$ & $0.00$ & $0.00$ & $1.94$ 
\end{tabular}
\end{ruledtabular}
\label{tab2}
\end{table}

\begin{table}  
\caption{The coordinates $\vec{r}_m=(x_m,y_m,z_m)$ for $m=1, \dots 10$
of the 10 particles forming a crystal cell,
with respect to the center of the cell, as obtained after 
an equilibration time $10^2 \Delta t$ 
at $T=0.03 U_A/k_B$ of the artificial crystal
(Fig.~\protect\ref{artificial}). The characteristic distances, with an
error on the last decimal digit, 
are: $l_1/a=0.53$, $l_2/a=0.59$, $l_3/a=0.07$, $l_4/a=1.43$, $l_5/a=0.05$,
$l_6/a=0.03$, $l_7/a=1.40$, $l_8/a=0.52$. The errors decrease as the
square root of the time of averaging.
For each particle $m$ of the cell, we denote with $n^{sc}_m$
the number of particles in the
crystal at a distance $r\leq b$ (in the soft-core) and with $n^{aw}_m$ 
the number of particles at a distance 
$b<r\leq c$ (in the attractive well).
}    
\begin{ruledtabular}
\begin{tabular}{c r r r l l }  
$m$ & $x_m$    & $y_m$    & $z_m$  & $n^{sc}_m$ & $n^{aw}_m$ \\ 
\hline
1   & $l_1$    & $l_2$    & $l_3$  & 6  & 15 \\
2   & $-l_1$   & $l_2$    & $-l_3$ & 6  & 15 \\
3   & $-l_1$   & $-l_2$   & $-l_3$ & 6  & 15 \\
4   & $l_1$    & $-l_2$   & $l_3$  & 6  & 15 \\
5   & $-l_4$   & $0$      & $-l_5$ & 3  & 24 \\
6   & $l_4$    & $0$      & $l_5$  & 3  & 24 \\
7   & $-l_6$   & $-l_7$   & $-l_8$ & 4  & 21 \\
8   & $l_6$    & $-l_7$   & $l_8$  & 4  & 21 \\
9   & $-l_6$   & $l_7$    & $-l_8$ & 4  & 21 \\
10  & $l_6$    & $l_7$    & $l_8$  & 4  & 21 
\end{tabular}
\end{ruledtabular}
\label{tab3}
\end{table}  

\begin{figure}
\Large{\bf (a)}\includegraphics[width=3cm,height=3cm,angle=0]{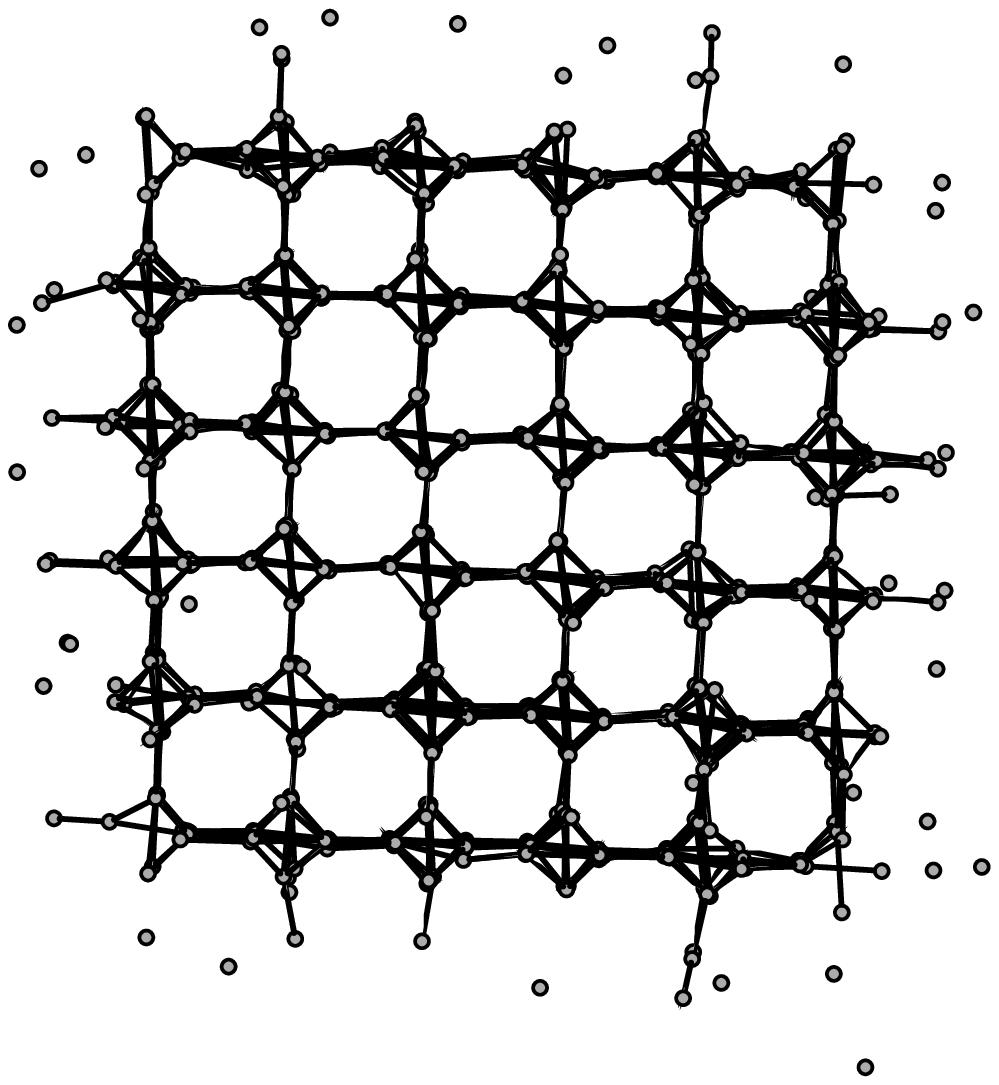} 
\Large{\bf (b)}\includegraphics[width=3cm,height=3cm,angle=0]{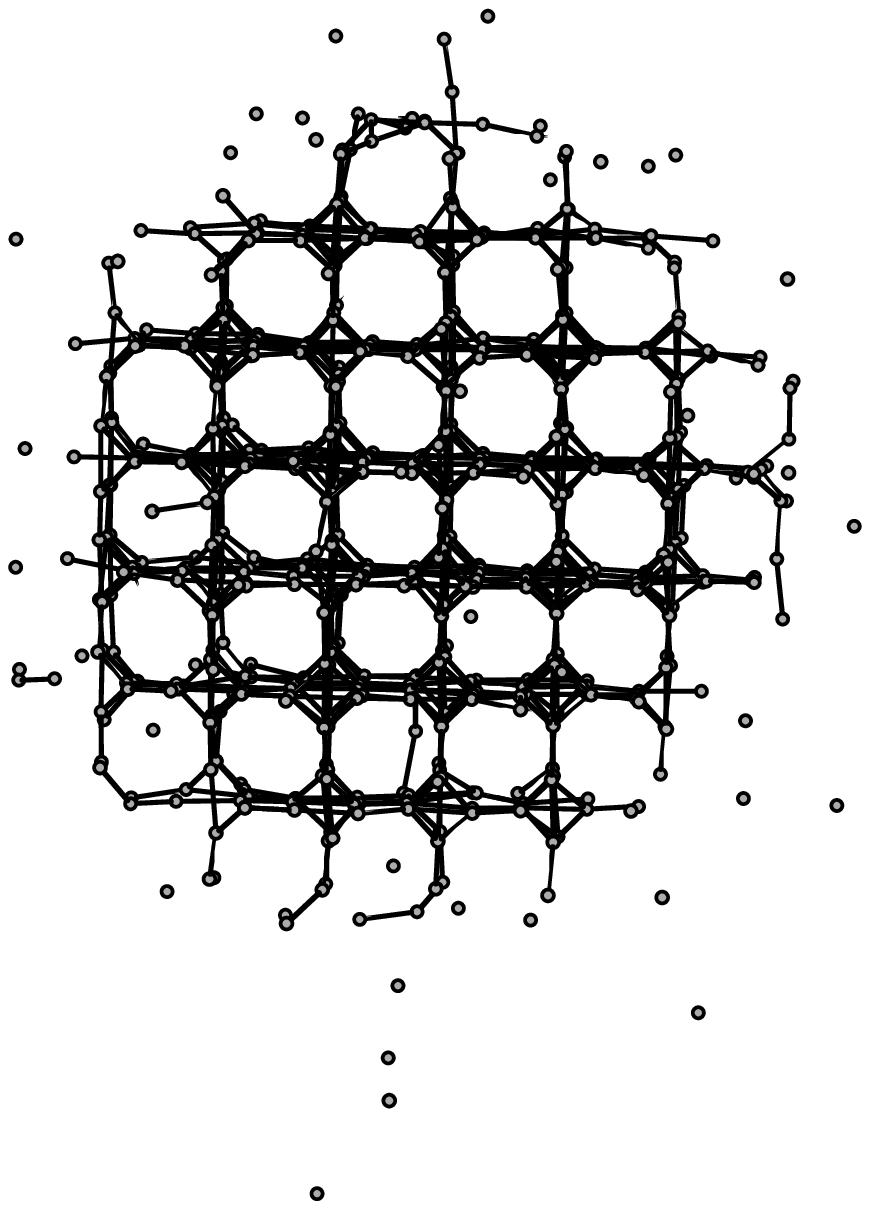} 
\caption{The artificial crystal configuration for $N=720$ particles 
equilibrated with a MD simulation at $T=0.10U_A/k_B$ (a) and at 
$T=0.52U_A/k_B$ (b). Bonds
connect particles at distance $r/a\leq 1.2$. The radius of the
particles is {\it not} in scale with the 
distances. The crystal seeds are in equilibrium with the gas phase and
show many defects.
The tilt present at the lower $T$ in (a), disappears at the
higher $T$ in (b).
} 
\label{higherT} 
\end{figure}

Surface effects could be responsible for the tilt that can be seen in
Fig.~\ref{artificial}d. In a system with $N=720$,
this tilt disappears when the sample is
equilibrated at higher $T$ (Fig.~\ref{higherT}).

\begin{figure}
\includegraphics[width=8cm,height=6.5cm,angle=0]{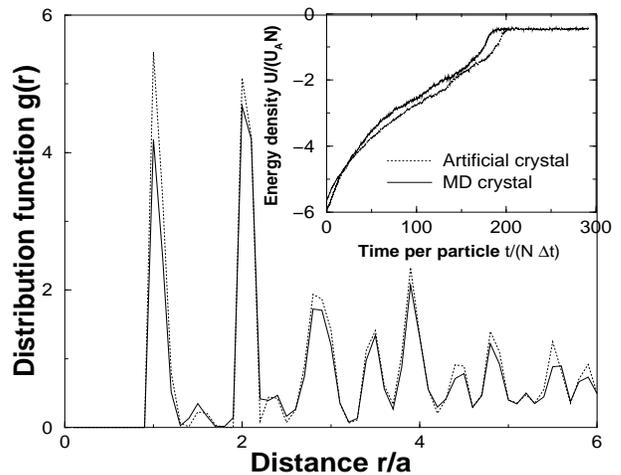} 
\caption{The radial distribution functions $g(r)$ for the MD crystal
(solid line)
and the artificial crystal (dotted line), 
both equilibrated at $T=0.48U_A/k_B$, are very close to each other. 
Inset:
the potential energy density $U/N$ 
for the MD crystal (solid line) with $N=490$
and the artificial crystal (dotted line) with $N=720$, 
both equilibrated at $T=0.60U_A/k_B$ starting from a configuration
equilibrated at $T=0.48U_A/k_B$, as function of the time divided by the 
number of particles $N$.} 
\label{comp_MD_art} 
\end{figure}

We compare the $g(r)$ (Fig.~\ref{comp_MD_art}) of the MD crystal in
Fig.~\ref{xtal} and of the artificial crystal in Fig.~\ref{artificial},
both equilibrated at $T=0.48U_A/k_B$.
Both functions show peaks located at the same distances, with
two large peaks at $r/a=1$ and $r/a=2$, consistent
with the presence of the two characteristic distances $a$ and $b$ in
the potential.
The comparison 
confirms that the proposed crystal is a good representation of the
crystal structure generated by the MD simulation. 
The slightly different intensities of the peaks of the 
$g(r)$ of the two systems are probably due to the defects of
the MD crystal. 

The validity of the artificial crystal as a good description of the real
crystal structure is confirmed also by the evolution of the 
potential energy per particle (inset in Fig.~\ref{comp_MD_art}) when 
the MD crystal and the artificial crystal are heated, from the configuration
equilibrated at $T=0.48U_A/k_B$, to $T=0.60U_A/k_B$. 
Both samples equilibrate to the same energy.
The starting
potential energy is, as expected, in both cases 
greater than the ground state energy $U_0/N=-8.45U_A$,
calculated from the number of
particles at distance $a\leq r< b$ and at distance $b\leq r < c$
(Table~\ref{tab3}), due to surface effects. We find analogous results
for the evolution of the kinetic energy.

\begin{figure}
\includegraphics[width=8cm,height=6.5cm,angle=0]{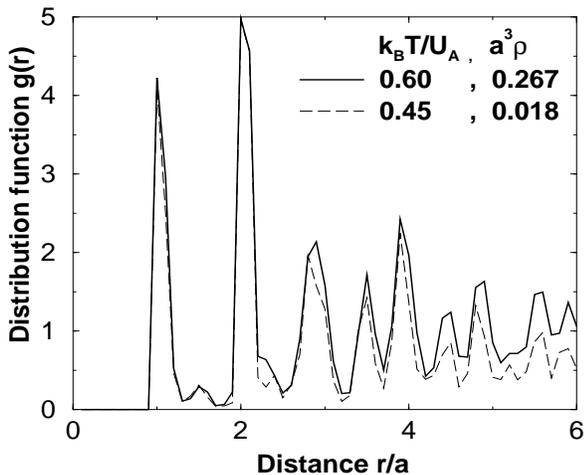}
\caption{Comparison of the radial distribution function $g(r)$ calculated
for two MD configurations obtained by cooling the system at different
densities:  the solid line is for the configuration at $k_BT/U_A=0.60$ and
$a^3\rho=0.267$, the dashed line for the configuration at $k_BT/U_A=0.45$
and $a^3\rho=0.018$.  The two functions
are very close to each other for distances $r/a\leq 4$,
showing that the crystal structure is the same at both densities. The
difference between the two functions is consistent with the presence of
defects and of the surrounding gas.} 
\label{2gr} 
\end{figure}

To test if the system has more than one crystal structure as function
of the density, we cool at $T=0.6U_A/k_B$ a fluid configuration
equilibrated at $T=0.8U_A/k_B$ and $\rho=0.267/a^3$, and compare the
resulting $g(r)$ with the case of the 
crystal seeds at $T=0.45U_A/k_B$ and $\rho=0.018/a^3$, finding no
relevant differences (Fig.~\ref{2gr}).  
At the same time, the attempt of finding alternative artificial
crystal structures has revealed, after an appropriate equilibration, 
that the only stable structure is the one presented in
Fig.~\ref{artificial}. 
We therefore assume that the system, at least for this
choice of the potential's parameters, has one single crystal structure
independent of $\rho$, within the considered range of densities. 

\begin{figure}
\includegraphics[width=7.5cm,height=6cm,angle=0]{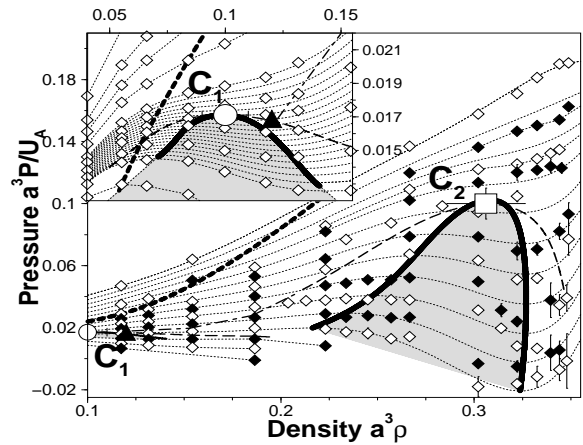}
\caption{The MD $P$-$\rho$ phase diagram.  
The thick dashed line is the gas-crystal first-order transition
line at the equilibrium. 
The calculations in the region below this line are for the
metastable fluid states.
Main panel:
the diamonds (full and open) 
are the MD calculations for (bottom to top) 
$k_B T/U_A = 0.570$, 0.580, 0.590, 0.600, 0.610, 0.620, 0.630, 0.640,
0.650, 0.660, 
0.670, 0.675, 0.685, 0.700. 
The dotted
lines are the isotherms calculated by polynomial interpolations of the
points at constant $T$ and, at the same time, of the points at constant
$\rho$.  
The circle at $\rho=0.1/a^3$ 
is the gas-LDL critical point $C_1$.  The square at $\rho=0.306/a^3$ is the
gas-HDL critical 
point $C_2$.  The solid thick lines connecting the local minima and maxima
along the isotherms are the spinodal lines associated to each critical
point and the shaded regions are the associated mechanically unstable
regions.  The dashed lines, passing through the critical points,
are the coexistence regions associated to
each critical point.  The meeting point of the gas-HDL coexistence line
with the gas-LDL
coexistence line gives the possible triple point (full triangle at
$\rho\simeq 0.12/a^3$).  Where
not shown, the errors are smaller then the symbols.  Inset: enlarged view
around $C_1$. Symbols are as in the main panel.  The diamonds are the MD
calculations for (bottom to top) $k_B T/U_A= 0.580$ 0.590, 0.595, 0.600,
0.603, 0.606, 0.609, 0.620, 0.630, 0.640, 0.660.} 
\label{MD} 
\end{figure}

Starting from a configuration with the crystal seed described above, we
equilibrate the system at different densities and temperatures.  We define
the system to be in the solid phase if, after a time
$10^6~(ma^2/U_A)^{1/2}\simeq 3\times 10^3 \Delta t$, the crystal seed is
growing, or we consider it in a fluid phase if the seed is melting. The
cases in which the trend is not clear within the simulation time are
considered as belonging to the first-order transition region
\cite{finite_size}.  The crystallization pressure rapidly increases with
$\rho$ and $T$, giving a first-order transition line (in the thermodynamic
limit) that separates the equilibrium $P$-$\rho$ phase diagram in a
high-$T$ fluid and a low-$T$ crystal (Fig.~\ref{MD}) \cite{nature}. 

\subsection{The supercooled liquids}

At equilibrium, there is no (stable) liquid phase.  A phase diagram
without equilibrium liquid phase is expected
\cite{Hagen,Frenkel-Lekkerkerker-PRL} for systems with an inter-particles
potential with a narrow attractive part, such as the one we are
considering here.  However, the liquid is present as a metastable
(supercooled) phase with respect to the crystal phase \cite{nature}.  To
study the metastable phase diagram, we equilibrate the system for each
$\rho$ from a gas configuration at $T=0.70U_A/k_B$ and then rapidly cool
it to the desired $T\geq 0.57U_A/k_B$, calculating $P$, $g(r)$ and the
total potential energy $U\equiv\sum_{i<j}^N U(|r_i-r_j|)$.

We find that the supercooled fluid phase has a life-time
longer than $3\times 10^3 \Delta t$ (the standard length of our
simulations) for  $\rho\lesssim 0.20/a^3$ at $T\approx 0.57
U_A/k_B$, for $\rho\lesssim 0.27 /a^3$ at $T\approx 0.65 U_A/k_B$ and
for $\rho\lesssim 0.34/a^3$ at $T\approx 0.70 U_A/k_B$.  
The system is equilibrated in the fluid phase for $t\geq 20 \Delta t$,
after which we average $P$, $g(r)$ and $U$ over the time. 
We calculate each state point by averaging 
the configurations for  $3\times 10^2 \Delta t \leq t \leq 3 \times
10^3 \Delta t$.  We estimate the errors by dividing the configurations
in 90 non-overlapping intervals of $30 \Delta t$, that we assume independent.

For larger $\rho$, the system spontaneously crystallizes (homogeneous
nucleation process). Thus, we only average over configurations that occur
before nucleation.  To be certain that our estimates are carried out in
the fluid phase, we study 
\begin{equation} 
S(\vec{q},t)\equiv \frac{1}{N} \left\langle \sum_{j,k}^N 
e^{i\vec{q}\cdot[\vec{r}_j(t)-\vec{r}_k(t)]}
\right\rangle ~ ,
\end{equation}
where $\vec{r}_j(t)$ is the position of particle $j$ at time $t$ and
$\vec{q}$ is the wavevector.  
At equilibrium, the average of $S(\vec{q},t)$, over
the time and the wavevectors with the same module,
is the structure factor $S(q)$, 
describing the spatial correlation in the system. 
Therefore, $S(\vec{q},t)$ describes the time-evolution of the
spatial correlation along the wavevector $\vec{q}$.
In particular, for a crystal-like configuration, with a long-range
order, there is at least one
wavevector such that $S(\vec{q},t) \sim O(N)$, while
for a fluid-like configuration is $S(\vec{q},t) \sim O(1)$ for
all $\vec{q}$. 

\begin{figure}
\includegraphics[width=8cm,height=6.5cm,angle=0]{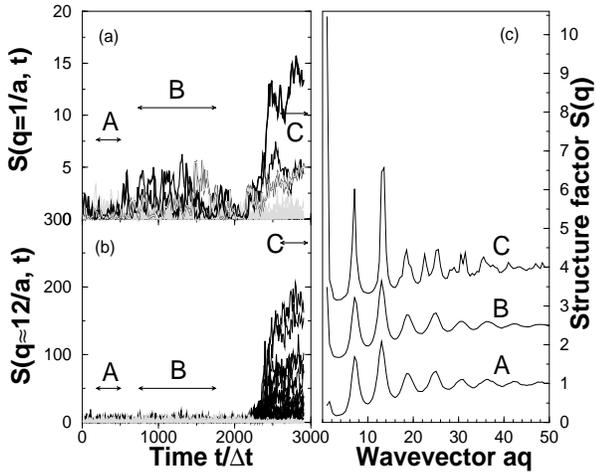}
\caption{MD calculations.  (a) Time evolution of the structure factor
$S(\vec{q},t)$ at $T=0.62 U_A/k_B$ and $\rho=0.27/a^3$, for wavevectors
with modulus $q= 1/a$ and for a time $t/\Delta t=3000$;
in the time interval ``A'' with 
$200\Delta t \leq t \leq 500\Delta t$,
it is $S(\vec{q},t)\sim O(1)$ for any
$\vec{q}$; in the time interval ``B'' with $700 \Delta t \leq t \leq
1800 \Delta t$, for 
six wavevectors there is an increase in $S(q=1/a,t)$;  in the interval
``C'' with $2100 \Delta t\leq t \leq 3000 \Delta t$, for the same six
wavevectors there 
is a larger increase.  
(b) As in (a) but for $q\approx 12/a\approx 4\pi/a$; in this case
there is a large increase in $S(q\approx 12/a,t)$, more then one order of
magnitude, only in the time interval ``C'' for several wavevectors,
revealing the formation of a crystal seed.  
(c) The structure factor $S(q)$, given by the average over the
dimensionless wavevectors $aq$ with the same modulus and the
average over the time intervals ``A'', ``B'' and ``C'' of
$S(\vec{q},t)$. 
The curves for ``B'' and ``C'' are offset by 1.5 and 3,
respectively. All the curves go to 1 for large $q$.
In the interval ``A'', $S(q)$ is liquid-like.  In the interval ``B'',
$S(q)$ is still liquid-like but with an increase for $q\rightarrow 0$,
 while in the interval ``C'' is solid-like, with two large peaks at
$q\approx 2\pi/(b/2)= 2\pi/a$ and $q\approx 2\pi/(a/2)$, corresponding
to the soft-core radius and the hard-core radius, respectively, and a
large value for $q\rightarrow 0$.} 
\label{Sofq} 
\end{figure}

The time-evolution of $S(\vec{q},t)$ for a typical simulation inside the
nucleation region is presented in Fig.~\ref{Sofq}.  To limit the
computational effort, we consider $9\times 10^4$ wavevectors with modulus
$q\le 100/a$, that is much larger than the wavevectors of
the largest peak of the crystal structure factor,
$q\approx 2\pi/(a/2)$, corresponding to the hard-core radius
(Fig.~\ref{Sofq}c). 
Three different regimes can be distinguished in the example in
Fig.~\ref{Sofq}. 

\begin{figure}
\includegraphics[width=8cm,height=6.5cm,angle=0]{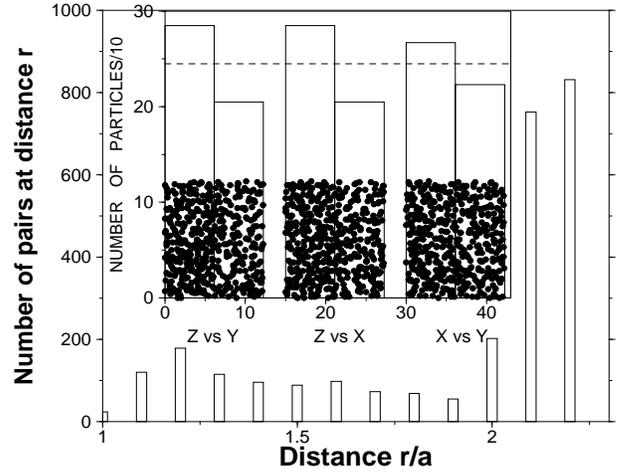}
\caption{Inset: projections of the 3D configuration of $N=490$ particles in a
system of size $V^{1/3}=12.25a$ and corresponding to the largest peak in
the time interval ``B'' in Fig.~\protect\ref{Sofq}a. The projections are
(from left to right) Z vs Y, Z vs X and X vs Y. The histograms of number
of particles as functions of the abscissa is over-imposed on each
projection. Projections and histograms are shifted for 
clarity. Each histogram bin corresponds to half of the box size.  The
dashed line shows the average number of particles in each bin for a
uniform configuration ($N/2=245$).  The largest deviation from the average
is $40 > \protect\sqrt{N} \approx 22$, i.e.  twice the statistic
fluctuation for a random distribution of particles.
Main panel: 
The number of pairs of particles at a relative distance
$r_i<r\leq r_{i+1}$ for the MD configuration in the inset,
with $r_0=0$, $r_1/a=1$ and $r_{i+1}-r_i=a/10$ for $i>1$. The histogram
shows a large maximum corresponding to the attractive range $2\leq
r/a<2.2$, a broad maximum around $r/a=1.2$ and a small number of pairs at
the hard-core distance $r=a$.  Therefore, the preferred relative distance
for pairs of particles within the soft-core range $1\leq r/a<2$ is, for
this configuration, $r/a\simeq 1.2$.} 
\label{sep}
\end{figure}

\begin{figure}
\includegraphics[width=8cm,height=8.5cm,angle=0]{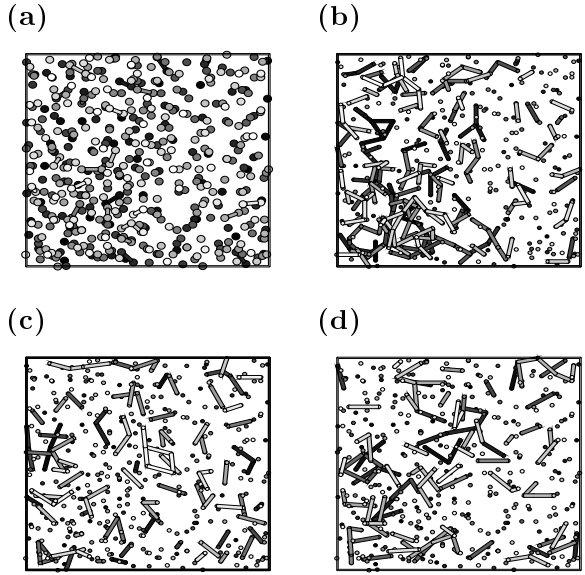}
\caption{Spatial distribution of pairs of particles at various distances
for the MD configuration shown in Fig.~\protect\ref{sep}.
The radius of the particles is {\it not} in scale with the distances.
In (a) the darker particles are farther away from the observation point.
Bonds connect particles at the hard-core distance $1\leq r/a< 1.1$ in
panel (a); 
at distance $1.1< r/a\leq 1.2$ in panel (b);  
at distance $1.3< r/a\leq 1.4$ in panel (c);
at distance $1.5< r/a\leq 1.6$ in panel (d).
The non-uniform distribution of bonds is clearly seen in panel (b).} 
\label{bonds}
\end{figure}

\begin{figure}
\includegraphics[width=8cm,height=7.5cm,angle=0]{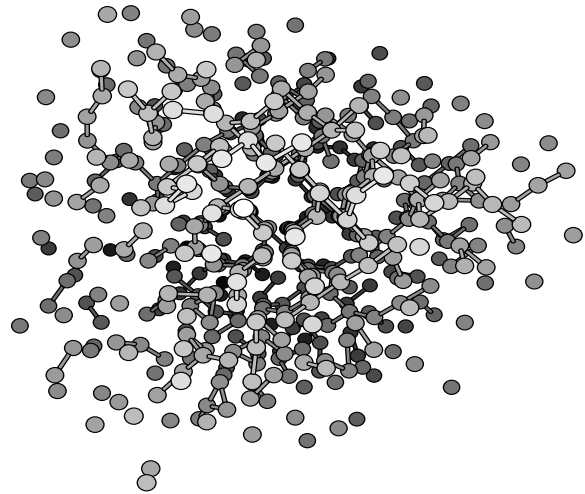}
\caption{The last MD configuration in the time series in
Fig.~\protect\ref{Sofq}. A crystal nucleus surrounded by gas is clearly
seen. Bonds connect particles at distance $r/a\leq 1.1$.  The 3D
perspective is given as in Fig.~\protect\ref{xtal}. The radius of
particles is {\it not} in scale with the distances.} 
\label{cryst}
\end{figure}

\begin{itemize} 
\item[(i)] A short-time regime ``A'', in which
$S(\vec{q},t)\sim O(1)$ for any $q$ ($q=1/a$ and 
$q\approx 12/a \approx 4\pi/a$  are shown
in Fig.~\ref{Sofq}a,b).  Averaged on this interval (curve A in
Fig.~\ref{Sofq}c) the $S(q)$ is fluid-like. 

\item[(ii)] An intermediate-time regime ``B'', in which $S(\vec{q},t)$
for $q=1/a$ has an increase, but has no increase for $q\approx 4\pi/a$. 
Averaged on this interval (curve B in Fig.~\ref{Sofq}c) the $S(q)$ is
fluid-like, but with an increase for $q\rightarrow 0$.  This increase
indicates an increase of $K_T$, according to the equation 
\begin{equation}
k_B T \rho K_T=\lim_{q\rightarrow 0} S(q) ~, 
\label{compress}
\end{equation} 
where we use the Eq.~(\ref{K_T}) and the definition
$S(\vec{q})\equiv 1+\rho \hat{h}(\vec{q})$.  The increase of $K_T$ is
associated with the phase separation into two fluids with different
densities. 

To help visualize the phase separation, in Fig.~\ref{sep} we show the
three planar projections of the 3D configuration corresponding to the
largest peak in the time interval ``B'' for $q=1/a$ (Fig.~\ref{Sofq}a). By
dividing the box into two equal parts, the histograms of the number of
particles in each part (Fig.~\ref{sep}) show a separation in density
approximately at half box-length, corresponding to $q=4\pi(\rho/N)^{1/3}
\approx 1/a$ for $\rho=0.27/a^3$ and $N=490$, in agreement with the peak at
$q=1/a$ for the curve B in Fig.~\ref{Sofq}c.  In each projection, it is
possible to see regions of high density and low density (Fig.~\ref{sep}).

To better quantify the phase separation occurring in the configuration in
Fig.~\ref{sep}, we present (main panel in Fig.~\ref{sep}) 
the histogram of the number
of pairs of particles at a relative distance $r_i < r \leq r_{i+1}$, where
$r_{i+1}-r_i=a/10$. The histogram has a broad maximum around the distance
$r/a=1.2$ in the soft-core range, showing that there 
exists a subset of pairs of
particles that are at a preferred distance $1.1<r/a\leq 1.2$.  This subset
is the HDL that has a non-uniform distribution over space
(Fig.~\ref{bonds}b), consistent with the phase separation.

\item[{(iii)}] A long-time regime ``C'', in which is $S(\vec{q},t)\sim
O(N)$ for $q\approx 4\pi/a$, revealing the crystal nucleation process. The
$S(q)$ averaged over this time interval (curve C in Fig.~\ref{Sofq}c) is
solid-like. In the same interval, $S(\vec{q},t)$ for $q=1/a$ has a large
increase, corresponding to the increase of $K_T$ ($S(q)$ increases for
$q\rightarrow 0$), which is consistent with the phase separation between
the fluid and the crystal. As an example, in Fig.~\ref{cryst} we show the
last configuration of the time series in Fig.~\ref{Sofq}, where the
crystal structure, already observed in Fig.~\ref{xtal},
is clearly seen. 

\end{itemize}

The example in Fig.~\ref{Sofq}b shows the formation of a high density
fluid phase within the time interval ``B'', followed by the nucleation of
the crystal phase. The onset of the nucleation is marked by a large
increase of $S(\vec{q},t)$ for all the wavevectors corresponding to the
peaks in the crystal $S(q)$ and by a large step-like decrease of energy. 

\subsection{The phase diagram and the finite size effect}

By repeating the analysis described above 
for all the simulations inside the region with
nucleation---and discarding the data corresponding to the formation of the
nucleus---it is possible to calculate the state points corresponding to
the metastable fluid phase. The phase diagram in Fig.~\ref{MD} is based on
averages over a total of $10^5$--$10^6$ configurations in the
fluid phase, accumulated in independent runs. 

For completeness we recall here the main features of the phase diagram in
Fig.~\ref{MD} and presented in Ref.~\cite{nature}.  The (mechanically
unstable) region at high $\rho$ for $T\lesssim 0.67 U_A/k_B$, where $P$
decreases for increasing $\rho$, denotes the coexistence of gas and HDL.
The unstable region at low $\rho$ for $T\lesssim 0.603 U_A/k_B$ (inset in
Fig.~\ref{MD})  denotes the coexistence of gas and LDL. The coexistence
lines are obtained by using the Maxwell construction of the equal areas
\cite{Debenedetti}, suggesting the presence of a gas-LDL-HDL triple point. 

By definition, the spinodal lines (limit of stability of each phase with
respect to the coexisting phase) meet the coexistence lines in a critical
point. Therefore, by interpolation we estimate the gas-LDL critical point
$C_1$ at $k_BT_1/U_A=0.603 \pm 0.003$, $a^3 \rho_1=0.10 \pm 0.01$, $a^3
P_1/U_A = 0.0171 \pm 0.0005$ and, the gas-HDL critical point $C_2$ at
$k_BT_2/U_A=0.665 \pm 0.005$, $a^3\rho_2=0.306 \pm 0.020$, $a^3 P_2/U_A =
0.10 \pm 0.01$.  These values are consistent with the linear
interpolations of the MD isotherms (Fig.~\ref{MD}). 

The phase diagram resulting from the MD calculations is, as expected, in
agreement with the time-dependent analysis of the structure factor
presented above. For example, the case presented in Fig.~\ref{Sofq}
corresponds to a state point inside the gas-HDL coexistence region at a
density higher then the crystal nucleation density for $T=0.62 U_A/k_B$.
The nucleation of the (metastable) HDL phase is thus followed by the
crystal nucleation. 

\begin{figure}
\includegraphics[width=8cm,height=6.5cm,angle=0]{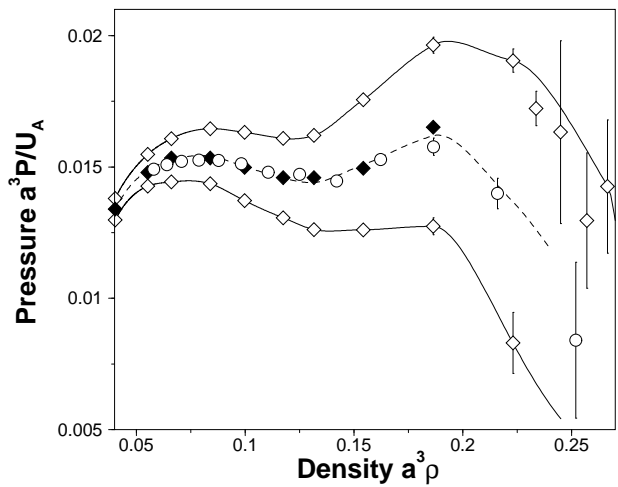}
\caption{Comparison between MD simulations for $N=490$ (full diamonds) and
$N=1728$ (circles) at $k_BT/U_A=0.595$. The results for the two sizes are
very close.
For comparison, we include also the calculations for
$N=490$ at $k_BT/U_A=0.60$ and 0.59 (upper open diamonds and lower open
diamonds, respectively) and the interpolation at $k_BT/U_A=0.595$ (dashed
line) between these two isotherms, showing the presence of two regions
with negatively sloped isotherms. The points calculated for $N=1728$ are 
also consistent with this interpolation, suggesting that the finite size
effect between $N=490$ and $N=1728$ is small.  Errors, where not shown,
are smaller than the symbols.} 
\label{compMD} 
\end{figure}

To estimate the finite size effect in our calculations, we compare the
results for $N=490$ and $N=1728$ for an isotherm below
both critical points (Fig.~\ref{compMD}).  
The calculations do not show any relevant finite
size effect, suggesting that the MD results for $N=490$ are reliable. 

\section{The radial distribution function analysis}
\label{HNCvsMD}

The interpretation of the HNC instability line is qualitatively consistent
with the MD spinodal line for the corresponding set of the potential's
parameters.  The projection of the MD spinodal line in the $T$-$\rho$
plane (not shown) has the same characteristics of the HNC instability
line, with two local maxima and one local minimum.  In both approaches,
the high-$\rho$ local maximum occurs at a temperature higher then the
temperature of the low-$\rho$ maximum and the presence of a triple point
is suggested by the presence of the local minimum. 

The quantitative HNC predictions for the locations of the two critical
points are, as expected, only partially consistent with the MD results. 
It is remarkable that the HNC estimates of 
the density of the low-$\rho$ local maximum ($\rho\approx 1/a^3$) 
and the temperature of the high-$\rho$ local maximum 
($T\approx 0.65 U_A/k_B$) of the instability line 
are close to the corresponding MD results for $C_1$ and $C_2$,
respectively. 

\begin{figure}
\includegraphics[width=8cm,height=6.5cm,angle=0]{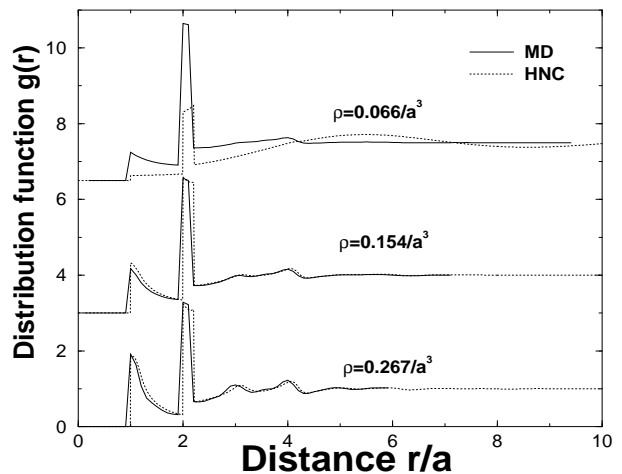}
\caption{Comparison between the $g(r)$ calculated in the HNC approximation
(dotted line) and by MD simulations (solid line). As an example, we
present the calculations at $T=0.64 U_A/k_B$ for densities (from top to
bottom)  $a^3\rho = 0.066$, 0.154, 0.267. For clarity a constant
value is added to the first two curves (6.5 and 3, respectively). The two
independent calculations are very close at intermediate densities.  At
large $r$ all the curves go to 1.} 
\label{hnc_md_gr} 
\end{figure}

An estimate of the agreement between the two methods can be evaluated by
comparing the calculations for $g(r)$ within the two approaches
(Fig.~\ref{hnc_md_gr}).  In contrast with what could be suggested by the
nature of the HNC approximation---i.e. the under-estimate of the indirect
correlation---the agreement is better at intermediate $\rho$ than at low
$\rho$ (Fig.~\ref{hnc_md_gr}). In particular, at low $\rho$ the HNC
approximation underestimates the probability of a particle penetrating the
soft-core or entering the attractive well.  At higher $\rho$, instead, the
estimates of the $g(r)$ within the two approaches are almost
indistinguishable. 

\begin{figure}
\includegraphics[width=8cm,height=6.5cm,angle=0]{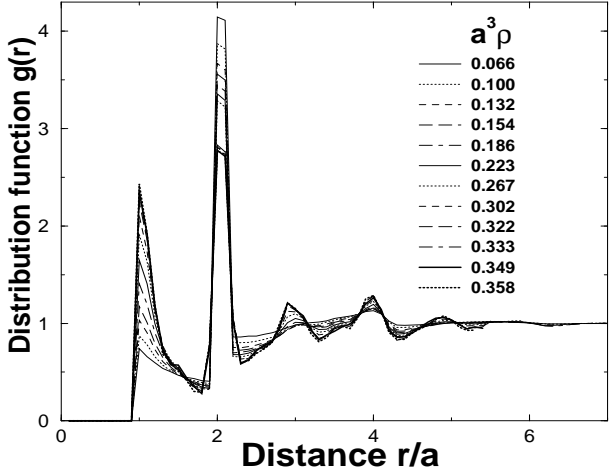}
\caption{The radial distribution function $g(r)$ calculated from the MD
results at $T=0.64U_A/k_B$ for densities $a^3\rho=0.066$, 0.100, 0.132,
0.154, 0.186, 0.223, 0.267, 0.302, 0.322, 0.333, 0.349, 0.358. With
increasing $\rho$, the peak at $r=a$ increases and the peak at $r=b$
decreases, while more peaks appear at larger $r$.} 
\label{md_gr}
\end{figure}

\begin{figure}
\includegraphics[width=8cm,height=6.5cm,angle=0]{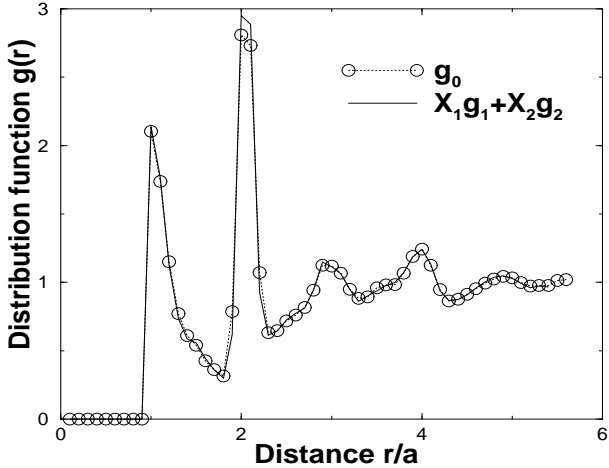}
\caption{The radial distribution function calculated from the MD
simulations at $T=0.64 U_A/k_B$ and $\rho=0.302/a^3$ (open circles) is
compared with the composition $X_1 g_1(r)+X_2 g_2(r)$ (solid line), where
$g_1(r)$ is the radial distribution function for the pure gas phase (at
$\rho=0.223/a^3$) and $g_2(r)$ is for the pure HDL phase (at
$\rho=0.349/a^3$) at the same $T$, with $X_1=0.3$ and $X_2=0.7$.}
\label{compose} 
\end{figure}

The $g(r)$ of the low-$\rho$ fluid is characterized by a large peak at $r=
b$ corresponding to the shortest attractive distance.  As a consequence of
the increase of the density, the peak at the hard-core distance $r=a$
increases while the peak at $r=b$ decreases, and additional peaks at
$r/a=3$, 4, ...  appear.  In Fig.~\ref{md_gr} we present the calculation
of the $g(r)$ for the gas phase, the gas-HDL coexisting region and the HDL
phase.  In particular, by combining the radial distribution functions
evaluated in each pure phase, we can estimate the composition of the mixed
phase. For example, at $T=0.64U_A/k_B$ the radial distribution function
calculated at $\rho_0=0.302/a^3$ is $g_0(r)\simeq X_1 g_1(r)+X_2 g_2(r)$,
where $g_1(r)$ and $g_2(r)$ are the radial distribution functions at the
same $T$ and at $\rho_1=0.223/a^3$ and $\rho_2=0.349/a^3$, respectively,
with $X_1=0.3$ and $X_2=0.7$ (Fig.~\ref{compose}), revealing that the
system is composed approximately by $30\%$ of gas and $70\%$ of HDL. 

\begin{figure}
\includegraphics[width=8cm,height=6.5cm,angle=0]{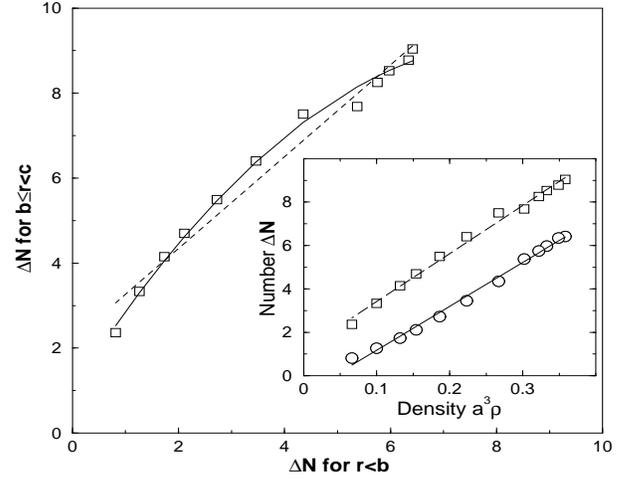}
\caption{Inset: the MD results at $T=0.64U_A/k_B$ 
for the cumulative number of particles $\Delta N$
within the repulsive range $r<b$ (circles) and within the attractive
range $b\leq r<c$ (squares), 
increasing linearly with $\rho$ in this range of densities;
the linear fit of the data gives a slope $20.2\pm 0.5$ for 
the solid line and a slope $22.1\pm 0.6$ for the 
dashed line. 
Main panel: the data in the inset plotted one versus the other
to show that within the mechanically unstable region
($\rho>0.267/a^3$) $\Delta N$ increases approximately in the same way with 
$\rho$ both within the attractive and within the repulsive range 
(the dashed line is a linear fit of all the data with slope 
$1.08\pm 0.06$) and increases faster within the attractive range 
at lower densities (the solid line is a quadratic fit).}
\label{md_nr}
\end{figure}

From the Eq.~(\ref{g_r}),
by using the $g(r)$ calculated from the MD simulations, we evaluate the
average number of particles $N(r)=\int dN(r)$ within a sphere of radius
$r$ (Fig.~\ref{md_nr}).  This analysis reveals that 
the number of particles $\Delta N$ within the repulsive range and 
within the attractive range increases linearly with $\rho$ (inset
Fig.~\ref{md_nr}) 
and that the increase is faster within the
attractive range (Fig.~\ref{md_nr}), for the densities we studied. 
In particular, the number of
particles within the attractive range $b\leq r<c$
increases from 2.5 to $9\ll \sum_{m=1}^{10} n^{aw}_m/10=19.2$ 
estimated for the artificial crystal
(Table~\ref{tab3}). 

\section{Absence of a density anomaly}

In Ref.~\cite{nature} it has been noted that the possibility of a second
fluid-fluid critical point is not necessarily restricted to systems with a
density anomaly, at least from a theoretical point of view. Here we
present the explicit thermodynamic calculations for this result. 

The defining relation for the density anomaly is given by
\begin{equation}
\left.\frac{\partial V}{\partial T}\right|_P<0 ~ ,
\label{tmd}
\end{equation}
or
\begin{equation}
\left.\frac{\partial S}{\partial V}\right|_T 
\left.\frac{\partial V}{\partial P}\right|_T > 0 ~ ,
\label{newtmd}
\end{equation}
for the Maxwell relation
\begin{equation}
\left.\frac{\partial V}{\partial T}\right|_P =
- \left.\frac{\partial S}{\partial P}\right|_T ~ ,
\end{equation}
where $S$ is the entropy.
Since 
\begin{equation}
\left.\frac{\partial V}{\partial P}\right|_T < 0
\end{equation}
holds for a mechanically stable phase,
Eq.~(\ref{newtmd}) can be rewritten as
\begin{equation}
\left.\frac{\partial S}{\partial V}\right|_T <0 ~ .
\label{SeV}
\end{equation}

From the differential expression of the thermodynamic potential at
constant $T$, we know that
\begin{equation}
TdS=dE+PdV ~ , 
\end{equation}
where $E\equiv U+K$ is the total energy, with $U$ and $K$
total potential and kinetic energy,
respectively.  
Therefore, it is
\begin{equation}
\left.\frac{\partial S}{\partial V}\right|_T = 
\frac{1}{T} \left.\frac{\partial U}{\partial V}\right|_T + \frac{P(V,T)}{T}
\end{equation}
at constant $T$
and we can rewrite the density anomaly condition in Eq.~(\ref{SeV}) as
\begin{equation}
\left.\frac{\partial U}{\partial V}\right|_T + P(V,T) < 0
\label{new}
\end{equation}
at constant $T$.

\begin{figure}
\includegraphics[width=8cm,height=6.5cm,angle=0]{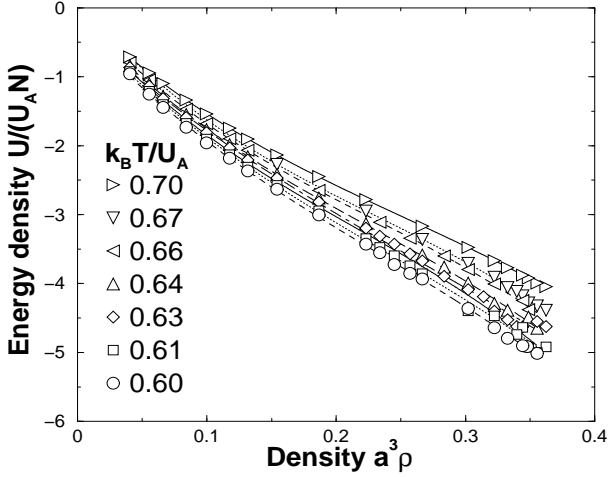}
\caption{The potential energy density $U/N$, calculated by MD simulations,
as a function of the density $\rho$ for temperatures (bottom to top)
$k_BT/U_A=0.60$, 0.61, 0.63, 0.64, 0.66, 0.67, 0.70. The symbols represent
the MD calculations, with errors smaller than the symbol's size. The lines
represent the cubic fit of the calculations with the parameters in
Table~\protect\ref{tab}.} 
\label{md_energy} 
\end{figure}

\begin{figure}
\includegraphics[width=8cm,height=6.5cm,angle=0]{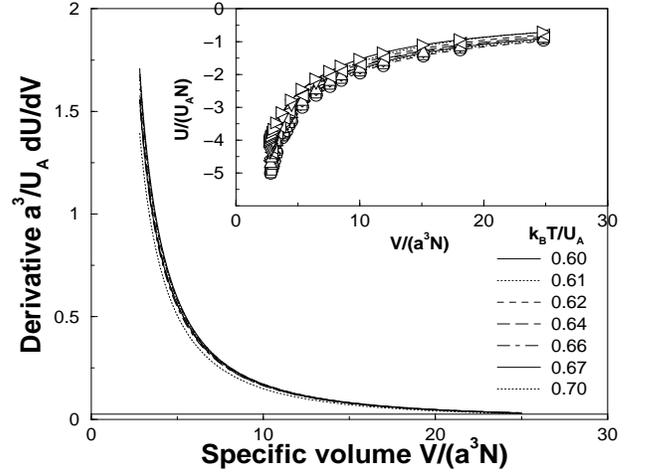}
\caption{The derivative $(\partial U/\partial V)_T$ calculated by using
the cubic expression in Fig.~\protect\ref{md_energy}, with the parameters
in Table~\protect\ref{tab}, 
as function of the specific volume $V/N$ for temperatures (top
to bottom) $k_BT/U_A=0.60$, 0.61, 0.63, 0.64, 0.66, 0.67, 0.70. The
derivative, in the considered range of $V/N$, is always larger than $0.025
U_A/a^3$ (bottom horizontal line).  Inset: The same MD results in
Fig.~\protect\ref{md_energy} for $U/N$ plotted as a function of $V/N$. The
symbols are the same as in Fig.~\protect\ref{md_energy}.
} 
\label{dUdV}
\end{figure}

To calculate the left-hand side of Eq.~(\ref{new}), we need to evaluate
$(\partial U/\partial V)_T$.  In Fig.~\ref{md_energy}, we show our MD
calculation for $U(\rho)$ at constant $T$. All the MD points can be fitted
with a third degree polynomial in $\rho$. The fitting parameters are given
in Table~\ref{tab} and are used to calculate the derivative $(\partial
U/\partial V)_T$, shown in Fig.~\ref{dUdV}.  Our calculations show a
potential energy $U$ increasing with $V$ (inset Fig.~\ref{dUdV}), with a
derivative always positive, thus wherever $P$ is positive, the condition
in Eq.~(\ref{new}) is not satisfied and there is no density anomaly. 

\begin{table}  
\caption{Parameters for the cubic fit
$U/N=a_0+a_1\rho+a_2\rho^2+a_3\rho^3$ of the MD calculations for
the potential energy density $U/N$ in 
Fig.~\protect\ref{md_energy} for
different 
temperatures. The errors on the fitting parameters are on the last
decimal digit.
}    
\begin{ruledtabular}
\begin{tabular}{c c c c c}  
$k_BT/U_A$ & $a_0$    & $a_1$    & $a_2$   & $a_3$ \\ 
\hline
0.60    & -0.2309  & -20.46   & 39.64   & -56.90 \\
0.61    & -0.2228  & -19.74   & 36.52   & -51.11 \\
0.63    & -0.1279  & -21.38   & 49.64   & -76.71 \\
0.64    & -0.1098  & -20.24   & 44.19   & -65.17 \\
0.66    & -0.0458  & -20.48   & 48.08   & -70.72 \\
0.67    & -0.0733  & -21.50   & 52.43   & -75.20 \\
0.70    & -0.0446  & -18.18   & 37.39   & -49.14 
\end{tabular}
\end{ruledtabular}
\label{tab}
\end{table}  

In the region where $P<0$ (at low $T$ and small $V$), the derivative $(\partial
U/\partial V)_T$ rapidly increases in such a way that Eq.~(\ref{new}) is
never satisfied.  Particularly, in the range of volumes considered, it is
always $(\partial U/\partial V)_T-0.025 U_A/a^3>0$, where $P=-0.025
U_A/a^3$ is the minimum pressure, reached for $T=0.6 U_A/k_B$ and
$V/N=3.31 a^3$ (Fig.~\ref{MD}). 
These results suggest that the density anomaly is ruled
out for this choice of parameters. At this stage is not clear if it is
ruled out for {\em any} choice of parameters for our potential in 3D (see
Ref.~\cite{PhysABig}). 

\section{Summary and conclusions}

We analyzed the phase diagram of a soft-core potential, similar to
potentials used in systems such as protein solutions, colloids, melts
and in pure systems such as liquid metals.
We use two independent numerical methods, integral equations in the HNC
approximation and MD simulations.
The comparison of the HNC results with previously-proposed soft-core
potentials suggests that the system has two fluid-fluid phase
transitions for an appropriate choice of parameters---energy and 
width---of the repulsive soft-core.
We select a set of potential parameters with a narrow
attractive well that gives a HNC instability line with two maxima and
suggests the presence of two critical points. 

The MD analysis shows, in agreement with the previous results for
potentials with a short range attraction \cite{Hagen}, a phase diagram
with no stable liquid phase. 
We analyze the crystal structure, characterized by the competition between
the attractive interaction at distance $r=b$ and the repulsive interaction
at $r=a<b$.
We show that the crystal, with 8-fold and 12-fold symmetries, is
independent on the density, within the considered range of densities. 

Hence, we study the metastable liquid phase, at
temperatures above and below the line of spontaneous crystal nucleation.
We find two liquids in the supercooled phase,
the LDL and the HDL, with two fluid-fluid transitions ending in two
critical points, the gas-LDL critical point $C_1$ and the gas-HDL
critical point $C_2$, as already shown in Ref.~\cite{nature}. 
Here we improve our estimate of the phase diagram and
we verify the absence of relevant finite size effects in the 
MD results. 

We compare these results with the HNC calculations, concluding
that the HNC approximation
underestimates the effect of the attractive interaction and
overestimates the effect of the repulsive interaction at low $\rho$, and
is in good agreement with the MD results at intermediate $\rho$.

Finally, by explicit calculation, we show that the condition for the
density anomaly is never satisfied in the range of $T$ and $V$ considered
here, as announced in Ref.~\cite{nature}. 
Our results suggest that the density anomaly is always ruled out for
this choice of potential parameters. 

In conclusion, the results of this paper
evoke an intriguing relation between
the absence of the density anomaly and the presence of a single
crystal phase, with higher density than the liquid phases,
in systems with two fluid-fluid phase transitions. This
relation, that deserves greater investigation, is consistent with the
fact that the substances with the density anomaly, and a hypothesized
second liquid-liquid critical point, have more than one crystal
structure, as in the case of water or carbon or silica.

\section{Acknowledgements}

We wish to thank M. C. Barbosa, P. V. Giaquinta, S. Mossa, G. Pellicane,
A. Scala, F. W. Starr and especially F. Sciortino for helpful suggestions
and for interesting and stimulating discussions.  We thank NSF Chemistry
Division (CHE-0096892) for support.


\begin{thebibliography}{99}
\bibitem{notaHNC} The equations of state can be obtained following three
different routes \cite{Hansen-McDonald}: by thermodynamic integration of
$K_T$, by using the Clausius virial theorem, or by integration of the
Maxwell relations between $P$, entropy and total energy. As already known
\cite{Hansen-McDonald}, due to the approximated nature of the HNC closure,
the results of the three different routes do not coincide, i.e. the theory
is thermodynamically inconsistent.  The inconsistency can be partially
removed by modifying the closure in such a way that it depends on a
parameter.  This parameter is usually fixed by requiring that the virial
pressure is equal to the pressure calculated through the compressibility
route. In this way a more accurate estimate of the spinodal line is
obtained and, due to the thermodynamic consistency, also the binodal (or
coexistence) line can be calculated.  However the numerical procedures
becomes much more computational-time consuming than the solution of the
HNC equations without the thermodynamic consistency. Therefore, for a
first extensive investigation in the parameter space of the model
potential, we choose to solve the integral equations in the HNC
approximation without imposing the thermodynamic consistency.
\bibitem{finite_size} It is known that in finite-systems a first-order
transition gives rise to a transition region rather than a transition line
(well defined only in the thermodynamic limit) \cite{Frenkelbook}.
\end{thebibliography}
\end{document}